\documentclass[%
 reprint,
superscriptaddress,
 amsmath,amssymb,
 aps,
 prl,
]{revtex4-2}
\usepackage{graphicx}
\usepackage{dcolumn}
\usepackage{bm}
\usepackage{epsfig}
\usepackage{amsmath}
\usepackage{subfigure}
\usepackage{verbatim}
\usepackage{amssymb}
\usepackage{array}
\usepackage{booktabs}

\begin{document}

\title{Quantum LiDAR with Frequency Modulated Continuous Wave}

\author{Ming-Da Huang}
\author{Zhan-Feng Jiang}
\email{jzf@szu.edu.cn}
\author{Hong-Yi Chen}
\address{State Key Laboratory of Radio Frequency Heterogeneous Integration, Shenzhen University, Shenzhen 518060, China}
\author{Ying Zuo}
\address{Shenzhen Institute for Quantum Science and Engineering, Southern University of Science and Technology, Shenzhen 518055, China}
\address{International Quantum Academy, Shenzhen 518048, China}
\address{Guangdong Provincial Key Laboratory of Quantum Science and Engineering, Southern University of Science and Technology, Shenzhen 518055, China}
\author{Xiao-Peng Hu}
\address{National Laboratory of Solid State Microstructures and College of Engineering and Applied Sciences, Nanjing University, Nanjing 210093, China}
\author{Hai-Dong Yuan}
\address{Department of Mechanical and Automation Engineering, The Chinese University of Hong Kong, Shatin, Hong Kong SAR, China}
\author{Li-Jian Zhang}
\address{National Laboratory of Solid State Microstructures and College of Engineering and Applied Sciences, Nanjing University, Nanjing 210093, China}
\author{Qi Qin}
\email{qi.qin@szu.edu.cn}
\address{State Key Laboratory of Radio Frequency Heterogeneous Integration, Shenzhen University, Shenzhen 518060, China}

\begin{abstract}

The range and speed of a moving object can be ascertained using the sensing technique known as light detection and ranging (LiDAR). It has recently been suggested that quantum LiDAR, which uses entangled states of light, can enhance the capabilities of LiDAR. Entangled pulsed light is used in prior quantum LiDAR approaches to assess both range and velocity at the same time using the pulses' time of flight and Doppler shift. The entangled pulsed light generation and detection, which are crucial for pulsed quantum LiDAR, are often inefficient. Here, we study a quantum LiDAR that operates on a frequency-modulated continuous wave (FMCW), as opposed to pulses.
We first outline the design of the quantum FMCW LiDAR using entangled frequency-modulated photons in a Mach-Zehnder interferometer, and we demonstrate how it can increase accuracy and resolution for range and velocity measurements by $\sqrt{n}$ and $n$, respectively, with $n$ entangled photons. We also demonstrate that quantum FMCW LiDAR may perform simultaneous measurements of the range and velocity without the need for quantum pulsed compression, which is necessary in pulsed quantum LiDAR. Since the generation of entangled photons is the only inefficient nonlinear optical process needed, the quantum FMCW LiDAR is better suited for practical implementations. Additionally, most measurements in the quantum FMCW LiDAR can be carried out electronically by down-converting optical signal to microwave region.

\end{abstract}
\maketitle

\begin{figure}[b]
\centering
\includegraphics[width=0.8\linewidth]{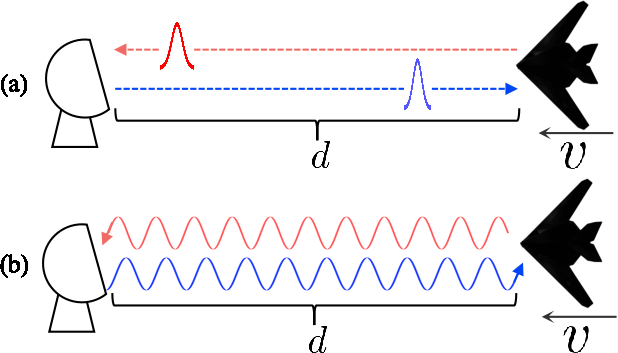}
\caption{The sketches of (a) pulsed LiDAR and (b) FMCW LiDAR. }
\label{fig:0}
\end{figure}

Several industries, such as automotive, robotics, unmanned aerial vehicles, etc., heavily rely on light detection and ranging (LiDAR) \cite{kimNanophotonicsLightDetection2021}. By delivering laser pulses and measuring the pulses that are reflected back from the target, as Fig. \ref{fig:0} (a) shown, pulsed LiDAR may measure distance and speed. The time it takes for each pulse to return to the sensor, or time of flight (ToF), can be used to calculate the range, and the Doppler shift of the returning pulses can be used to calculate the relative velocity of the object.

With the emergence of quantum metrology \cite{giovannettiQuantumMetrology2006}, the quantum version of LiDAR has been proposed that can achieve better precision and resolution \cite{giovannettiQuantumenhancedPositioningClock2001,giovannettiPositioningClockSynchronization2002,macconeQuantumRadar2020,lloydEnhancedSensitivityPhotodetection2008,changQuantumenhancedNoiseRadar2019,barzanjehMicrowaveQuantumIllumination2015,zhuangQuantumRangingGaussian2021,liuEnhancingLIDARPerformance2019,reichertQuantumenhancedDopplerLidar2022,shapiroQuantumPulseCompression2007,zhuangUltimateAccuracyLimit2022,zhuangEntanglementenhancedLidarsSimultaneous2017,huangQuantumLimitedEstimationRange2021,blakey2022quantum}. 
The earlier ideas concentrate on quantum-enhanced pulsed LiDAR, which substitutes entangled light for the conventional pulses.
Using entangled pulsed light with a high time-bandwidth product, quantum-enhanced pulsed LiDAR has demonstrated the ability to estimate range and velocity simultaneously through the application of quantum pulsed compression \cite{shapiroQuantumPulseCompression2007}. However, these quantum-enhanced pulsed LiDAR systems \cite{zhuangEntanglementenhancedLidarsSimultaneous2017,huangQuantumLimitedEstimationRange2021} typically involve inefficient nonlinear optical processes (in addition to the generation of entangled pulses) and require a lossless propagation channel. The systems are further complicated by the difficulty of monitoring the Doppler shift of the entangled pulses \cite{avenhausFiberassistedSinglephotonSpectrograph2009,thekkadathMeasuringJointSpectral2022}.

In addition to pulsed LiDAR, frequency modulated continuous wave (FMCW) technology can also be used to perform LiDAR operations, as Fig. \ref{fig:0} (b) shown. In contrast to pulsed LiDAR, FMCW LiDAR may receive signals constantly, eliminating the delay between pulses and eliminating the need for a minimum detection distance  \cite{liAdvancedDriverAssistance2022}.  By coherently combining echo and reference light to create a beat signal, FMCW LiDAR can estimate range and velocity at the same time \cite{zhengOpticalFrequencymodulatedContinuouswave2005,riemensbergerMassivelyParallelCoherent2020,lukashchukDualChirpedMicrocomb2022}. Moreover, FMCW LiDAR is more resistant to background noise \cite{qianVideorateHighprecisionTimefrequency2022,hsuFreeSpaceApplicationsSilicon2022} and better suited for on-chip integration \cite{behroozpourLidarSystemArchitectures2017,zhangLargescaleMicroelectromechanicalsystemsbasedSilicon2022,lihachevLownoiseFrequencyagilePhotonic2022}.

We study a quantum LiDAR with FMCW in this letter, which has three main contributions. We first present a scheme for the quantum-enhanced FMCW LiDAR, which is fundamentally different from the quantum pulsed LiDAR. This scheme involves a quantum FMCW light field of $n$-entangled photons with frequency modulation in a Mach-Zehnder (MZ) interferometer. By utilizing the entanglement among both the time and path, it improves the precision limit by $\sqrt{n}$ and the resolution by $n$ over the classical counterpart for the measurement of range and velocity. Second, using triangle frequency modulation, the quantum FMCW LiDAR enables the decoupling of range and velocity measurements. This makes it possible to estimate the target's position and speed at the same time without using a quantum pulsed compression system like a quantum pulsed LiDAR. Finally, the quantum FMCW LiDAR system is better suited for practical implementation. In addition to the ability of the quantum FMCW LiDAR's beat signal to down-convert the optical frequency enabling a more precise and compact electronic measurement of Doppler shift, the only inefficient nonlinear optical process that the system uses are the creation of the entangled photons. As a result, the system is easily applicable to integrated photonic platforms, such as thin film lithium niobite \cite{zhuIntegratedPhotonicsThinfilm2021,jinOnChipGenerationManipulation2014,xueUltrabrightMultiplexedEnergyTimeEntangled2021}, which can effectively generate, manipulate, and detect quantum signals.

We first present the basic working principle of the FMCW LiDAR, then extend it to the quantum FMCW LiDAR. 

\begin{figure}[t]
\centering
\includegraphics[width=1\linewidth]{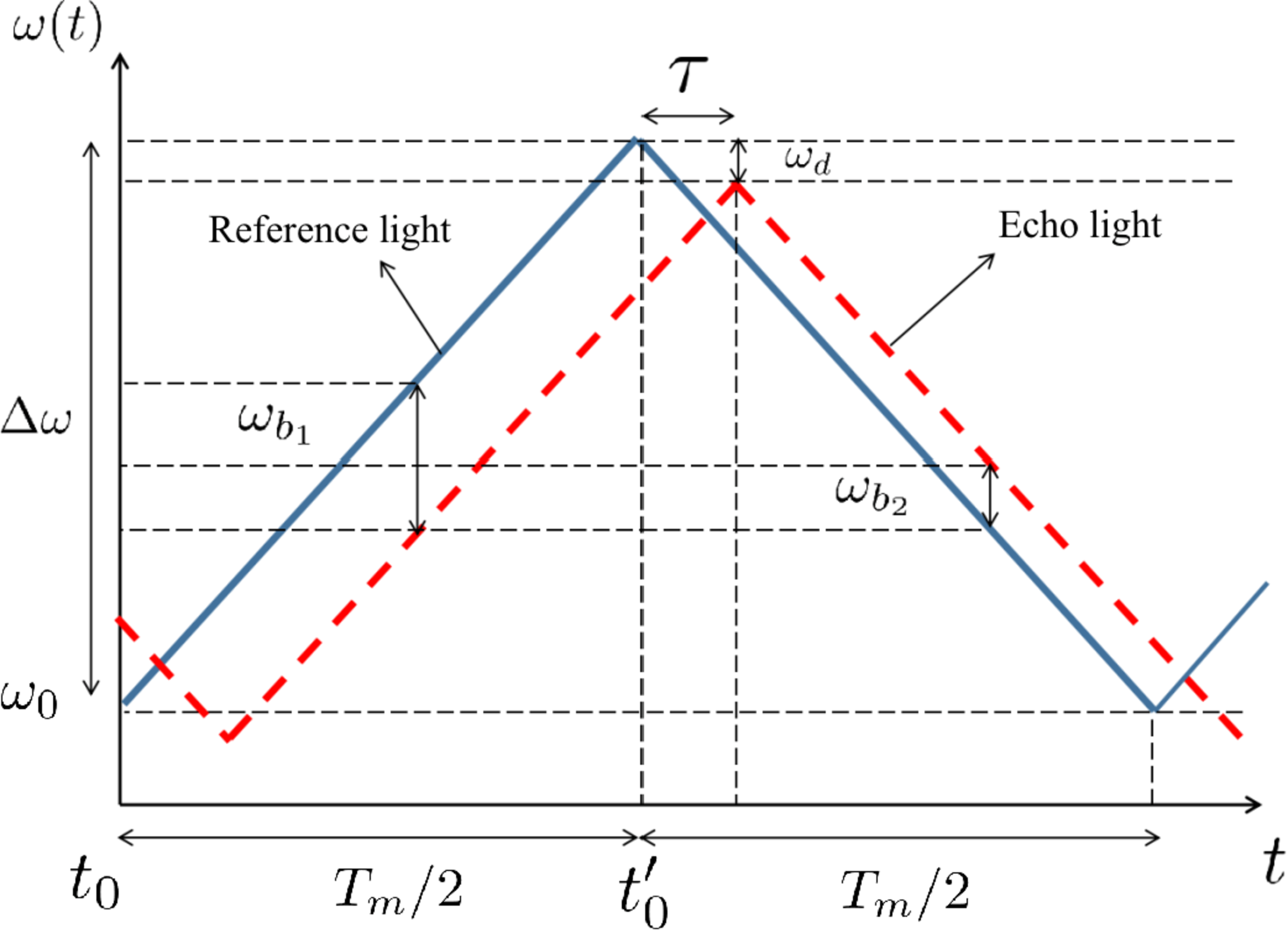}
\caption{The variation of angular frequency $\omega(t)$ of the reference light and the echo light under the triangle frequency modulation with an initial modulation time $t_{0}=t'_{0}-T_{m}/2$. For the reference light, $\omega(t)=\frac{\Delta \omega}{T_{m} / 2}\left(t-t'_{0}\right)+(\omega_{0}+\Delta \omega)$ in rising edge with a initial angular frequency $\omega_{0}$ and $\omega(t)=-\frac{\Delta \omega}{T_{m} / 2}\left(t-t'_{0}\right)+(\omega_{0}+\Delta \omega)$ in falling edge with a initial angular frequency $\omega_{0}+\Delta\omega$. $T_{m}$ and $\Delta\omega$ are modulation period and bandwidth respectively}
\label{fig:1}
\end{figure}

The FMCW LiDAR measures the range and velocity of the object by sending frequency modulated continuous wave and detecting the echo light that bounces back from the object \cite{zhengOpticalFrequencymodulatedContinuouswave2005}. The echo light has a time delay, denoted as $\tau$, which is related to the target's distance ($d$), and a Doppler shift ($\omega_d$), which is related to the object's relative radial velocity ($v$). The velocity and range can then be determined by contrasting the reference light with the echo light. A widely used frequency modulation is the triangle frequency modulation, where the frequency of the continuous wave first increases, then decreases linearly, as shown in Fig.\ref{fig:1}. 
In this case the time of flight and the Doppler shift can be obtained as
\begin{equation}\label{equ2}
\begin{aligned}
\tau&=\frac{2d}{c}=\frac{T_{m} }{2\Delta \omega }\frac{\omega_{b_{1}}+|\omega_{b_{2}}|}{2},\\
\omega_{d}&= \omega_{c}\frac{2v}{c}=\frac{\omega_{b_{1}}-|\omega_{b_{2}}|}{2} ,
\end{aligned}
\end{equation}
here $T_m$ is the period of modulation, $c>>v$ is the speed of light, and $\omega_{c}$ is the center frequency of the FMCW light field, $\omega_{b_{1}}$ and $\omega_{b_{2}}$ are the frequency differences between the reference light and the echo light at the rising and falling edges respectively, as shown in Fig. \ref{fig:1}. $\omega_{b_1}$ and $\omega_{b_2}$, which are the frequencies of the beat signal produced by mixing the reference light and the echo light, can then be used to determine the range and velocity. 

The phase of the beat signal also depends on the relative distance as $\theta\propto\omega_{c}\tau=2\pi d/\lambda_{c}$ (see Sec. IB of the supplemental material), where $\lambda_{c}$ is the center wavelength of the FMCW light field. We include the phase of the beat signal, $\theta_{0(1)}$ in the rising (falling) edge, in the parameters to be estimated in addition to the beat frequencies to account for the effects on the precision for the estimation of frequencies in the presence of the unknown phases. It's important to keep in mind that $\theta_{0(1)}$ depends on the initial time selections made for the rising and falling edges. The phase changes by $2\pi$ when the echo light travels a distance of one wavelength, making it highly sensitive to the relative distance. This property can be used to make detailed contours of the target's surface by measuring a set of $\theta$ from different points on the surface, similar to interferometric imaging with monochromatic light \cite{turbideAlloptronicSyntheticAperture2012,terrouxSyntheticApertureLidar2017}.

In classical optics, an optical FMCW beam can be described as an oscillating electromagnetic field $E_{\omega_{c}}(r,t)$ as
\begin{equation}
 E_{\omega_{c}}(r,t)\propto E_{\omega_{c}}(t)=\alpha e^{i \int_{t_{0}}^{t} \omega(t) d t},
\end{equation}
where $\alpha$, $\omega(t)$ and $\omega_{c}$ are the amplitude, modulating frequency and the center frequency respectively. Here for simplicity we consider a single spatial and polarization mode. When the field is modulated periodically, it can be decomposed into discrete frequencies with a spectrum $s_{\omega_{c}}(\omega_{j})=\frac{1}{T_{m}} \int_{t_{0}}^{t_{0}+T_{m}}d t E_{\omega_{c}}(t) e^{-i  \omega_{j} t} $, $\omega_{j}=2\pi j/T_{m}$ with $j\in\{0,1,2,\ldots,\infty\}$. 

In quantum optics, the classical light can be described by a coherent light $|\alpha(\omega)\rangle$. The frequency modulation transforms the coherent state to a FMCW coherent state $\mathop{\otimes}_{j}\left|\alpha s_{\omega_{c}}^{*}\left(\omega_{j}\right)\right\rangle$ with \begin{equation}
\hat{a}^{+}(\omega) \rightarrow \sum^{\infty}_{j=0} s_{\omega_{c}}^{*}\left(\omega_{j}\right) \hat{a}^{+}\left(\omega_{j}\right),
\end{equation}
here $s(\omega_{j})$ is a normalized complex spectrum with $\sum_{j}|s_{\omega_{c}}^{*}\left(\omega_{j}\right)|^{2}=1$ \cite{capmanyQuantumModelElectrooptical2010,capmanyQuantumModellingElectrooptic2011}.
Under the slowly-varying envelope approximation ($\omega_{c}>>\Delta\omega$) \cite{tsangQuantumTheoryOptical2008}, the classical FMCW light field $E_{\omega_{c}}(t)$ can be described by a multi-mode coherent state $\mathop{\otimes}_{j}\left|\alpha s_{\omega_{c}}^{*}\left(\omega_{j}\right)\right\rangle$ 
with
\begin{equation}\label{equ1}
\hat{E}^{(+)}(t) \propto \sum^{\infty}_{j=0}\hat{a}(\omega_{j}) e^{-i \omega_{j} t},
\end{equation}
where $\hat{a}(\omega_{j})$ and $\hat{a}^{+}(\omega_{j})$ are annihilation and creation operator. 

We use finite bandwidth and discrete time approximations to model the quantum FMCW field in the time domain. In these approximations, a period of modulation, $T_m$, is divided into $j_B$ intervals with $j_B+1$ discrete time points. The annihilation operator at time $t_p$, the $p$-th discrete time point, is given by
\begin{equation}\label{equ8}
\hat{a}(t_{p}) \equiv \frac{1}{\sqrt{j_{B}+1}} \sum_{j=j_{c}-\frac{j_{B}}{2}}^{j_{c}+\frac{j_{B}}{2}} \hat{a}\left(\omega_{j}\right) e^{-i \omega_{j} t_{p}},
\end{equation}
where $\omega_{j_B} = 2\pi j_B/T_m$ is the finite bandwidth with $\omega_{j_B} \gg \Delta\omega$, $\omega_{j_c}$ represents a frequency closest to the central frequency of the FMCW light fields and $t_p = pT_m/j_B$ where $p \in \{p_0, p_0+1, p_0+2,\dots,p_0+j_B\}$ with the initial time point $p_0 \in \mathbb{Z}$. Due to periodicity, we have $\hat{a}(t_{p}+T_m)=\hat{a}(t_{p})$.
A Fock state in the time domain can then be written as
\begin{equation}
\left|n_{t_{p}}\right\rangle=\frac{1}{\sqrt{n_{t_{p}} !}}\left[\hat{a}^{+}(t_{p})\right]^{n_{t_{p}}}|0\rangle,
\end{equation}
where $n_{t_{p}}$ is the photon number at $t_{p}$. As $[\hat{a}(t_{p}), \hat{a}^{\dagger}(t_{p'})]=\delta_{p,p'}$, the band-limited Hilbert space can be spanned by the discrete time mode $I=\otimes_{t_{p}} \sum_{n_{t_{p}}}\left|n_{t_{p}}\right\rangle\left\langle n_{t_{p}}\right|=\sum_{\mathbf{n}}\left|\mathbf{n}\right\rangle\left\langle \mathbf{n}\right|$, where $|\mathbf{n}\rangle=|n_{t_{p_{0}}},\ldots,n_{t_{{p_{0}+j_{B}}}}\rangle$ is a photon number state within a period.

Under the finite bandwidth and discrete time approximation, the single photon state of FMCW in the frequency domain, $\sum_{j=0}^{\infty} s_{\omega_{c}}^{*}\left(\omega_{j}\right) \hat{a}^{+}\left(\omega_{j}\right)|0\rangle$, can be transformed to the time domain as
\begin{equation}\label{equ6}
|\psi_{1}\rangle\approx  \frac{1}{\sqrt{j_{B}+1}} \sum_{p=p_{0}}^{p_{0}+j_{B}} E_{\omega_{c}}^{*}\left(t_{p}\right) \hat{a}^{+}\left(t_{p}\right)|0\rangle ,
\end{equation}
where $E_{\omega_{c}}\left(t_{p}\right)$ is the FMCW field given in Eq. (\ref{equ1}) represented in the discrete time domain, $\omega_{c}$ is the center angular frequency, and $j_{B}$ is taken sufficiently large to include most frequency bands of the FMCW field. It can be seen that the single FMCW photon state in the time domain is typically a superposition of states at different times.
For a two-photon anti-frequency-correlation entangled state \cite{jinOnChipGenerationManipulation2014,xueUltrabrightMultiplexedEnergyTimeEntangled2021,giovannettiExtendedPhasematchingConditions2002} with a Gaussian correlation spectrum $G(\omega)$ with standard deviation $\sigma$, given by $\int d \omega G(\omega) \hat{a}_{1}^{+}\left(\omega_{c}+\omega\right) \hat{a}_{2}^{+}\left(\omega_{c}-\omega\right)|0\rangle$, the state of FMCW in the time domain is
\begin{equation}\label{equ7}
\begin{aligned}
&|\psi_{2}\rangle\approx\frac{1}{j_{B}+1} \sum^{p_{0}+j_{B}}_{p,q=p_{0}} \tilde{G}\left(t_{p}-t_{q}\right) E_{\omega_{c}}^{}\left(t_{p}\right) E_{\omega_{c}}^{}\left(t_{q}\right) \times\\
&\hspace{5cm}\hat{a}_{1}^{+}\left(t_{p}\right) \hat{a}_{2}^{+}\left(t_{q}\right)|0\rangle,
\end{aligned}
\end{equation}
where $\tilde{G}\left(t\right)$ is the Fourier transformation of $G(\omega)$. We note that the phases of the frequency-modulated biphoton are correlated at different time points, known as chirp entanglement. The preparation of $|\psi_{2}\rangle$ is discussed in Sec. V of the supplemental material.

The quantum FMCW LiDAR is achieved by using part of the entangled state as the signal light and the other part as the reference light. A beating signal is then obtained from the interference between the echo light, that is reflected back from the target, and the reference light. 

In general the $n$-photon FMCW NOON state can be used, where the signal can be represented as
\begin{equation}\label{equ4}
\begin{aligned}
\left|\psi_{n}(\vec{x})\right\rangle=\frac{1}{\sqrt{2 T_{m}}} \int^{t_{0}+T_{m}}_{t_{0}} d t&\left\{E_{\omega_{c}}^{* n}(t)\left|n_{t}, 0\right\rangle-\right.\\
&\left.E_{\omega_{c}-\omega_{d}}^{* n}(t-\tau) \left|0,n_{t}\right\rangle\right\},
\end{aligned}
\end{equation}
where the time delay $\tau$ and the Doppler shift $\omega_{d}$ are encoded in the echo light, $\vec{x}$ denotes the collections of the parameters to be estimated. To extract the information from $\left|\psi_{n}(\boldsymbol{O})\right\rangle$, an interferometric detection strategy  
\begin{equation}\label{equ5}
\hat{A}_{D}(t)=-(|n_{t}, 0\rangle\langle 0, n_{t}|+| 0, n_{t}\rangle\langle n_{t}, 0|),
\end{equation}
can be applied at each time when photons are detected. This detection strategy keeps track of the probability of projecting on the two eigenstates of $\hat{A}_{D}(t)$, which forms the quantum beating signal.

\begin{figure}[t]
\centering
\includegraphics[width=1\linewidth]{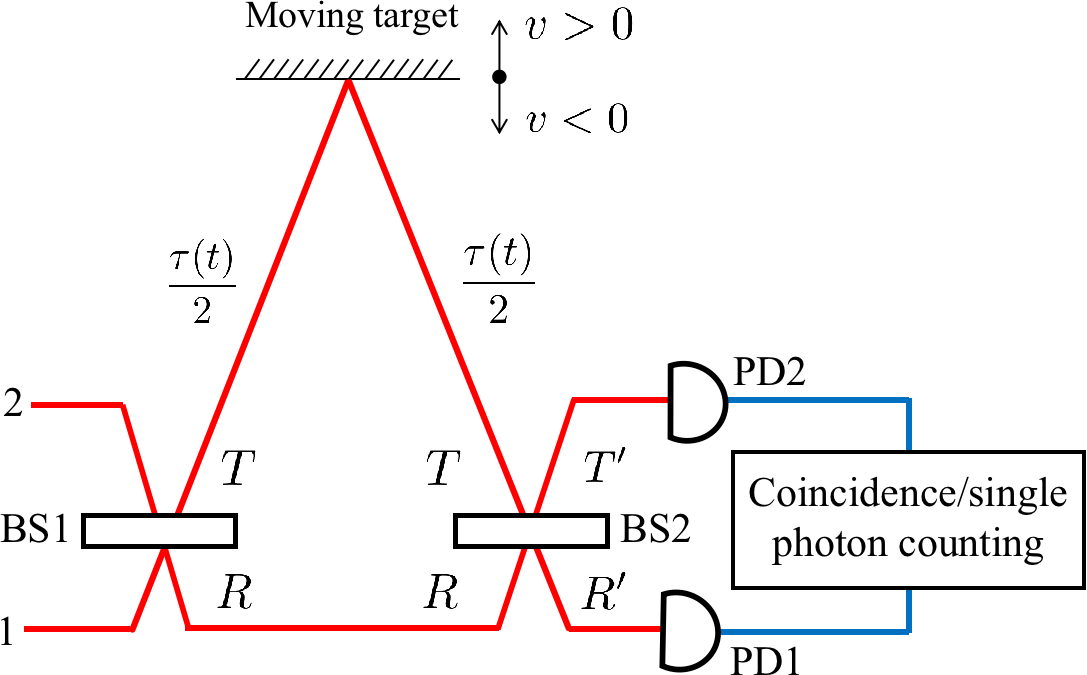}
\caption{The sketches of target ranging and velocity measurement by using MZ interferometer. The BS1 (BS2) and the PD1 (PD2) are the 50:50 beam splitter and the photon detector respectively. If the FMCW single photon state $|\psi_{1}\rangle$ input from port 1, the single photon counting is applied. If the FMCW biphoton entangled state $|\psi_{2}\rangle$ are input from port 1 and port 2 simultaneously, the coincidence photon counting is applied. The target velocity $v > 0$ ($v < 0$) when the target is moving away (toward) from the photon detectors which creates a red (blue) shift.}
\label{fig:2}
\end{figure}

The schematic physical implementation of quantum FMCW LiDAR scheme, for $n=1$ (for comparison) and $n=2$, is shown in Fig. \ref{fig:2}. It is similar to the MZ interferometer but with two differences. First, the target is always been placed in the middle of one arm to create specular-like reflection that simulates the round-trip propagation of the echo light. Second, the evolution of the light field is described with the frequency-independent time of light ($\tau(t)\approx\tau+(\omega_{d}/\omega_{c})t=2(d+vt)/c$ (if $|v|<<c$)) \cite{ivanovElaboratedSignalModel2020} instead of the frequency-dependent phase differences that is typically used in the MZ interferometer. The details of the derivation of the evolution and detection can be found in Sec. IIC and Sec. IID of the supplemental material.

A scheme of using bi-photon NOON state, $|\psi_{2}\rangle$, is sketched in Fig. \ref{fig:2}. 
After the BS1 and the free evolution, the state $|\psi_{2}\rangle$ becomes
\begin{equation}\label{equ10}
\begin{aligned}
&|\psi_{2}(\vec{x})\rangle=\frac{1}{2\left(n_{B}+1\right)}  \sum^{p_{0}+j_{B}}_{p, q=p_{0}} \tilde{G}\left(t_{p}-t_{q}\right)\times \\
&\left\{E_{\omega_{c}}^{*}\left(t_{p}\right) E_{\omega_{c}}^{*}\left(t_{q}\right) \hat{a}_{R}^{+}\left(t_{p}\right) \hat{a}_{R}^{+}\left(t_{q}\right)-\right.\\ 
&\enspace\left.E_{\omega_{c}-\omega_{d}}^{*}\left(t_{p}-\tau\right) E_{\omega_{c}-\omega_{d}}^{*}\left(t_{q}-\tau\right) \hat{a}_{T}^{+}\left(t_{p}\right) \hat{a}_{T}^{+}\left(t_{q}\right)\right\}|0\rangle
\end{aligned}
\end{equation}
showing both path and chirp entanglements.

At the detection side, two photon counters are used to keep track of the variation of joint detection probability with time. This is the second order quantum correlation function 
\begin{equation}
\langle\psi_{2}(\vec{x})|\hat{E}_{R^{\prime}}^{(-)}\left(t_{1}\right) \hat{E}_{T^{\prime}}^{(-)}\left(t_{2}\right) \hat{E}_{T^{\prime}}^{(+)}\left(t_{2}\right) \hat{E}_{R^{\prime}}^{(+)}\left(t_{1}\right)| \psi_{2}(\vec{x})\rangle
\end{equation}
with a time delay $\tau'=t_{1}-t_{2}$. Given that the time delay is normally distributed as shown in Eq. (\ref{equ10}), the probability of coincidence counting becomes 
\begin{equation}\label{equ14}
\begin{aligned}
p_{2}\left(1 \mid t, \vec{x}\right)=\int d \tau^{\prime}&\langle\psi_{2}(\vec{x})|\hat{E}_{R^{\prime}}^{(-)}\left(t\right) \hat{E}_{T^{\prime}}^{(-)}\left(t-\tau^{\prime}\right)\times\\
& \hat{E}_{T^{\prime}}^{(+)}\left(t-\tau^{\prime}\right) \hat{E}_{R^{\prime}}^{(+)}\left(t\right)| \psi_{2}(\vec{x})\rangle.
\end{aligned}
\end{equation}
Here, one of the photon is detected by PD1 at time $t_1=t$ while the other photon is detected by PD2 with an arbitrary time delay. So the probability of both photons detected by the same detector (PD1 or PD2) is $p_{2}\left(0 \mid t, \vec{x}\right)=1-p_{2}\left(1 \mid t, \vec{x}\right)$.

For triangular modulation, the scheme yields a piece-wise detection probability distribution
\begin{equation}\label{equ9}
\begin{aligned}
& p_n\left(1 \mid d, v, \theta_l, t\right) \approx \\
& \quad \frac{1}{2}+\frac{1}{2}\cos \left\{n\left[(-1)^{l}\frac{\Delta \omega}{T_m / 2} \frac{2 d}{c}+\omega_c \frac{2 v}{c}\right] t+ n\theta_l\right\}\end{aligned},
\end{equation}
where $l=0$ for $t\in[t_{d_{0}}-T_{m}/2, t_{d_{0}}]$, and $l=1$ for $t\in[t_{d_{1}},t_{d_{1}}+T_{m}/2]$. Here $n=1,2$ stand for the single photon and the two-photon entanglement states respectively. These detection probabilities oscillate with time, which form the quantum beat signals. The target distance $d$ and velocity $v$ can then be derived from the frequencies of the beat signal. 

The precision limit for the estimation of $\vec{x}=\{d,v,\theta_{0},\theta_{1}\}$ can be quantified by the Cramer-Rao bound (CRB) with $Cov(\hat{x})\geq \frac{1}{\nu}F_C^{-1}[\vec{x}]$, where $Cov(\hat{x})$ is the covariance matrix for unbiased estimator, $\hat{x}$, and $F_C[\vec{x}]$ is the classical Fisher information (CFI) matrix, $\nu$ is the number of times the procedure is repeated. In a period of FMCW modulation, the instaneous $F _{C}[\vec{x},t]$ varies over time $t$. To get overall CFI, the contribution of CFI in the whole period should be considered, e.g., $F _{C}[\vec{x}]=\int^{t_{d_{0}}}_{t_{d_{0}}-T_{m}/2}dt F _{C}[\vec{x},t]+\int^{t_{d_{1}}+T_{m}/2}_{t_{d_{1}}}dt F _{C}[\vec{x},t]$. Further, in order to exclude the covariance between initial phase $\theta_{0,1}$ and the beat frequencies resulted from different $t_{d_{0}}$ and $t_{d_{1}}$ (see Sec. IIIB of the supplemental material) \cite{rifeSingleToneParameter1974}, it is set that $t_{d_{0}}=-t_{d_{1}}=T_{m}/4$.

We can explicitly calculate $F_{C}[\boldsymbol{O}]$ and obtain the CRB as
\begin{equation}\label{equ15}
\begin{aligned}
\operatorname{Cov}(\hat{x})&\geq \frac{1}{\nu}F^{-1}_{C}[\vec{x}]=\\ \frac{1}{\nu T_m}
&\begin{pmatrix}
\frac{3}{n^{2}} \frac{c^{2}}{\Delta \omega^{2}}  & 0 & 0 & 0 \\ \\
0 & \frac{12}{n^{2}} \frac{c^{2}}{\omega_{c}^{2} T_m^{2}} & 0 & 0 \\ \\
0 & 0 & \frac{2}{n^{2}}  & 0 \\ \\
0 & 0 & 0 &  \frac{2}{n^{2}}
\end{pmatrix}.
\end{aligned}
\end{equation}
With the same output power, the covariance matrix with entangled photon pairs ($n=2$) is half of the single photon case ($n=1$). 

In the quantum FMCW LiDAR scheme, the resolution of $\{d,v\}$ is rooted in the discrete Fourier transform (DFT) of the measured quantum beating signal. The DFT corresponds to the maximum-likelihood (ML) estimators for $\vec{x}$ \cite{rifeSingleToneParameter1974, erkmenMaximumlikelihoodEstimationFrequencymodulated2013}. The classical CRB can be achieved with this ML estimator \cite{rifeSingleToneParameter1974}. Hence, the resolutions of $\{d,v\}$ is governed by the frequency resolution in the DFT. For the triangular modulation, if we perform the DFT independently on the beating signals, given in Eq. (\ref{equ9}), at each half period, a frequency resolution of $\frac{2\pi}{T_{m}/2}$ can be achieved. This leads to the resolutions of $\{d,v\}$ as (see Sec.IIID of the supplemental material for detail):
\begin{equation}
\Delta d=\frac{2 \pi}{ n} \frac{c}{2 \Delta \omega}, \quad \Delta v=\frac{2 \pi}{ n} \frac{c}{\omega_c  T_{m} }.
\end{equation}
This shows an improvement on the resolution of $\{d,v\}$ by a factor of 2 when two-photon entangled states are used to replace the coherent states. In addition, the initial phase of the quantum beat signals are multiplied by two indicating that a phase resolution of $\{\theta_{0},\theta_{1}\}$ is improved by a factor of 2 \cite{botoQuantumInterferometricOptical2000}. Similar analysis can be extended to the scenario with $n$-photon entangled states with an improvement of a factor of $n$.

The quantum FMCW LiDAR that we studied in this letter is distinct from the conventional quantum pulsed LiDAR.
A design of this scheme with entangled photon pairs ($n=2$) is illustrated with an MZ interferometer. The ability of the quantum FMCW LiDAR to detect range and velocity simultaneously with high precision is demonstrated. This is in contrast to the pulsed quantum LiDAR, which requires quantum pulsed compression.

Quantum FMCW LiDAR provides three main advantages over pulsed quantum LiDAR. First, unlike the quantum pulsed LiDAR, the quantum FMCW LiDAR does not require additional nonlinear optical processes to entangle or de-entangle photons in order to measure ranging and velocity simultaneously \cite{ zhuangEntanglementenhancedLidarsSimultaneous2017,huangQuantumLimitedEstimationRange2021}. Note that in order to produce the entangled state of light, both types of quantum LiDARs require nonlinear optical processes (such as spontaneous parametric down conversion). However, further nonlinear optical processes are needed for the quantum pulsed LiDAR to de-entangle the entangled pulsed light in order to measure the ToF and the Doppler shift separately. These nonlinear optical processes are also utilized to entangle various types of pulsed light in quantum pulsed LiDAR proposals \cite{zhuangEntanglementenhancedLidarsSimultaneous2017}, in order to produce entangled pulsed light with high time-bandwidth products. These nonlinear optical processes \cite{couteau2018spontaneous}, however, are typically inefficient (see Sec. VI of the supplemental material). The quantum FMCW LiDAR is able to use quantum resources more efficiently, which is beneficial for key performance metrics, such as signal-to-noise ratio, measurement range, etc.

Second, the beat signals, which are normally in the microwave range and can be precisely measured by cutting-edge electronic circuitry, are used in the quantum FMCW LiDAR to extract the information. The operational frequency band is downconverted in this way, greatly increasing the scheme's viability. Comparatively, the exact measurement of the time and the optical Doppler shift required by the quantum pulsed LiDAR can be challenging \cite{avenhausFiberassistedSinglephotonSpectrograph2009,thekkadathMeasuringJointSpectral2022}.

Lastly, a significantly more compact architecture is made possible by the quantum FMCW LiDAR's better suitability for on-chip integration. Since the quantum FMCW LiDAR's peak power can be substantially lower than that of the pulsed LiDAR, it is more suited for integrated photonics. In fact, the beam splitter, modulation, and entangled light source of the on-chip quantum FMCW LiDAR have all been successfully demonstrated \cite{jinOnChipGenerationManipulation2014,xueUltrabrightMultiplexedEnergyTimeEntangled2021} on the lithium niobite chip platform \cite{zhuIntegratedPhotonicsThinfilm2021} with comparatively high efficiency. 

The quantum FMCW LiDAR has a lot of benefits for practical implementations due to its energy efficiency, capability to analyze signals in the microwave spectrum, and appropriateness for on-chip integration. The approach can also be applied to other non-classical light fields \cite{plickParityDetectionQuantum2010,kacprowiczExperimentalQuantumenhancedEstimation2010,sahotaQuantumenhancedPhaseEstimation2013,sahotaQuantumCorrelationsOptical2015,larsonSupersensitiveAncillabasedAdaptive2017,zhangNonlinearPhaseEstimation2019,alodjantsEnhancedNonlinearQuantum2022,yuQuantumEnhancedOptical2020,zhangImprovingPhaseEstimation2021,gattoHeisenberglimitedEstimationRobust2022}, such as the multimode squeezed-vacuum state \cite{reichertQuantumenhancedDopplerLidar2022}, the photon-added coherent state \cite{agarwalNonclassicalPropertiesStates1991,francisPhotonaddedCoherentStates2020} and the optimal photon number state \cite{dornerOptimalQuantumPhase2009}.

\section*{ACKNOWLEDGMENTS}

\bibliography{bibtex}

\end{document}


\title{Supplemental Material of ``Quantum LiDAR with Frequency Modulated Continuous Wave"}

\author{Ming-Da Huang}
\author{Zhan-Feng Jiang}
\email{jzf@szu.edu.cn}
\author{Hong-Yi Chen}
\address{State Key Laboratory of Radio Frequency Heterogeneous Integration, Shenzhen University, Shenzhen 518060, China}
\author{Ying Zuo}
\address{Shenzhen Institute for Quantum Science and Engineering, Southern University of Science and Technology, Shenzhen 518055, China}
\address{International Quantum Academy, Shenzhen 518048, China}
\address{Guangdong Provincial Key Laboratory of Quantum Science and Engineering, Southern University of Science and Technology, Shenzhen 518055, China}
\author{Xiao-Peng Hu}
\address{National Laboratory of Solid State Microstructures and College of Engineering and Applied Sciences, Nanjing University, Nanjing 210093, China}
\author{Hai-Dong Yuan}
\address{Department of Mechanical and Automation Engineering, The Chinese University of Hong Kong, Shatin, Hong Kong SAR, China}
\author{Li-Jian Zhang}
\address{National Laboratory of Solid State Microstructures and College of Engineering and Applied Sciences, Nanjing University, Nanjing 210093, China}
\author{Qi Qin}
\email{qi.qin@szu.edu.cn}
\address{State Key Laboratory of Radio Frequency Heterogeneous Integration, Shenzhen University, Shenzhen 518060, China}

\maketitle

\section{The basic principle of FMCW Lidar}

\subsection{Measurement of the target ranging and velocity in FMCW Lidar}

\begin{figure}
\centering
\subfigure[]{
\includegraphics[width=0.4\linewidth]{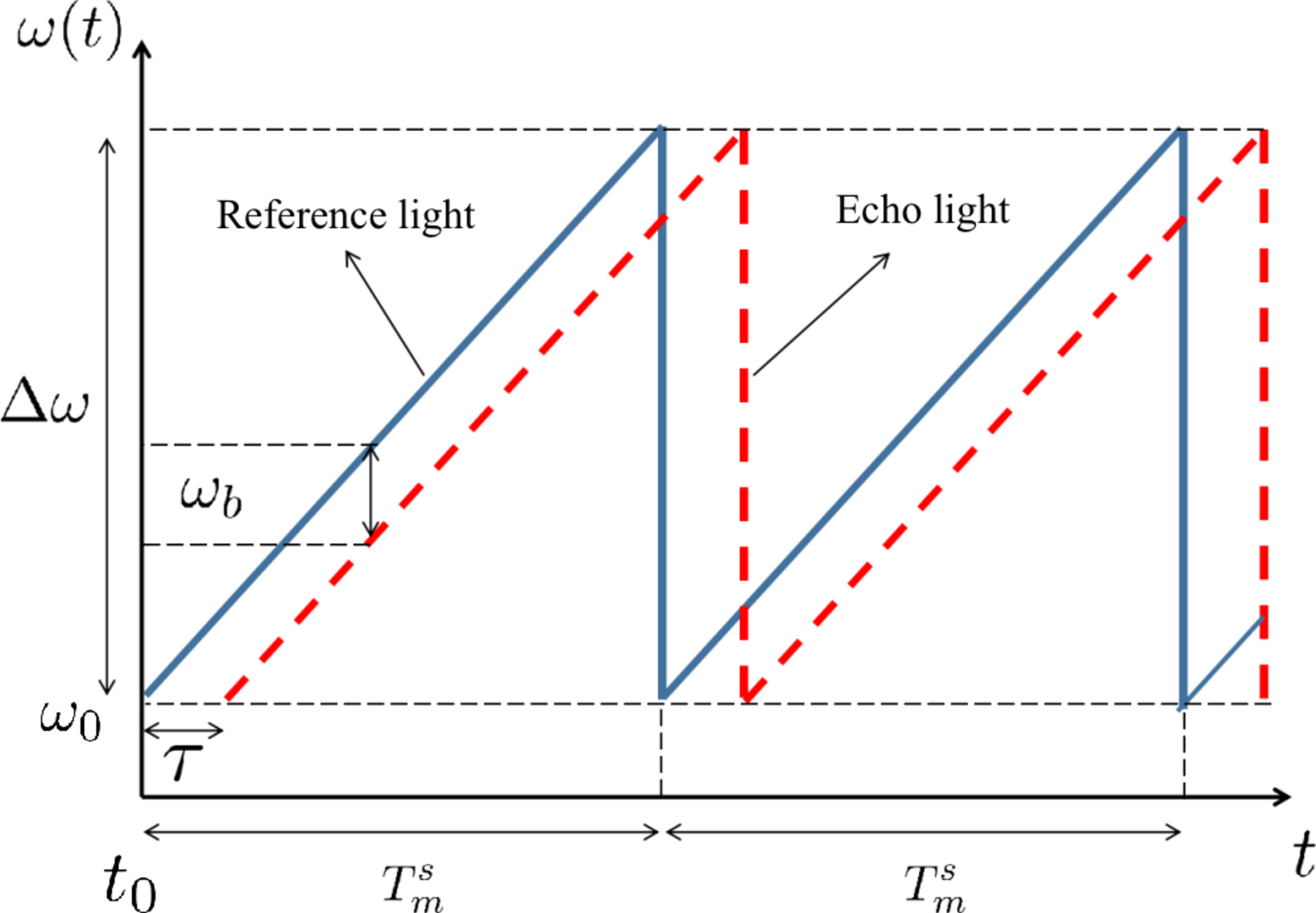}}
\hspace{0.01\linewidth}
\subfigure[]{
\includegraphics[width=0.4\linewidth]{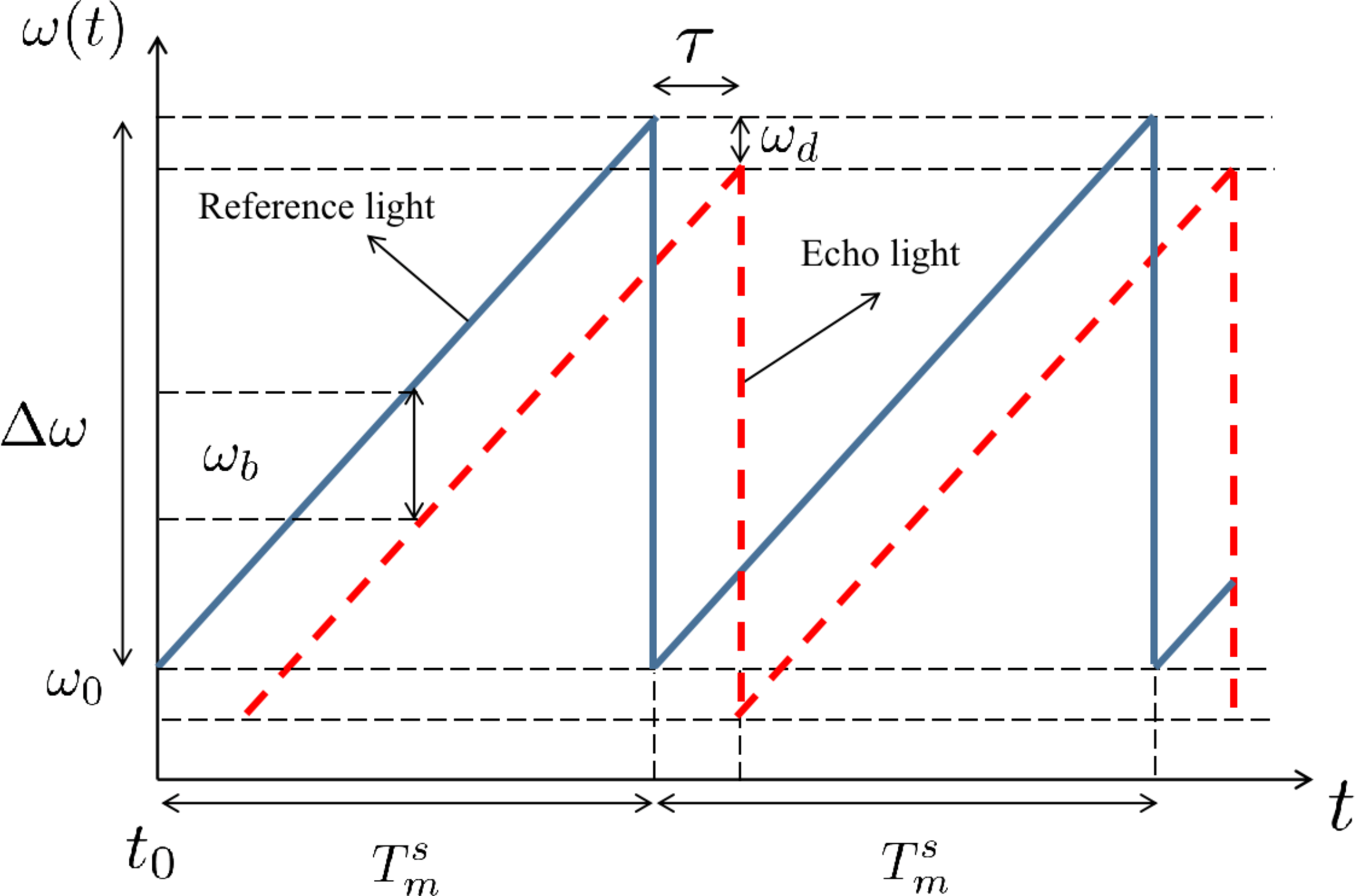}}
\vfill
\subfigure[]{
\includegraphics[width=0.4\linewidth]{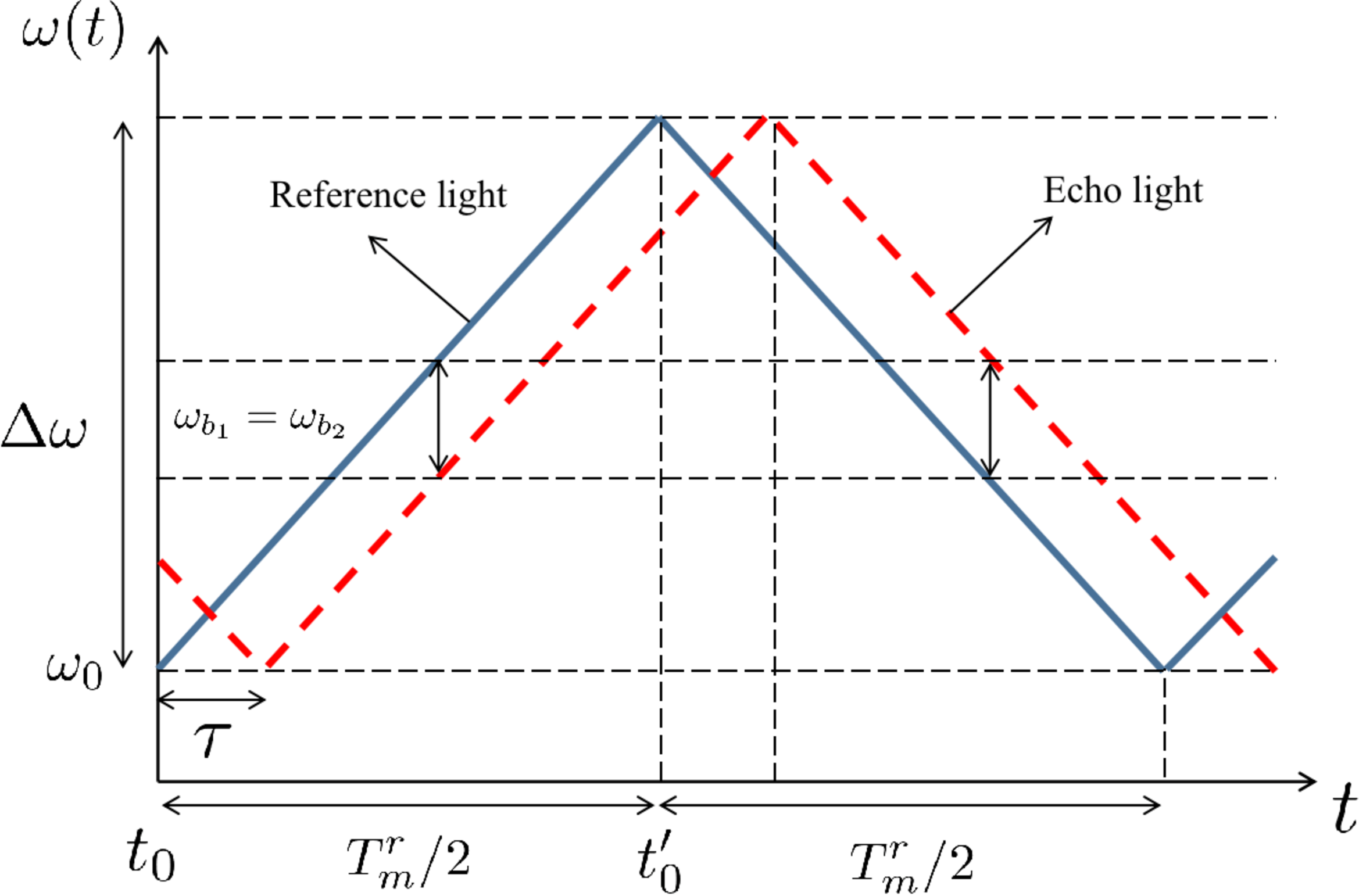}}
\hspace{0.01\linewidth}
\subfigure[]{
\includegraphics[width=0.4\linewidth]{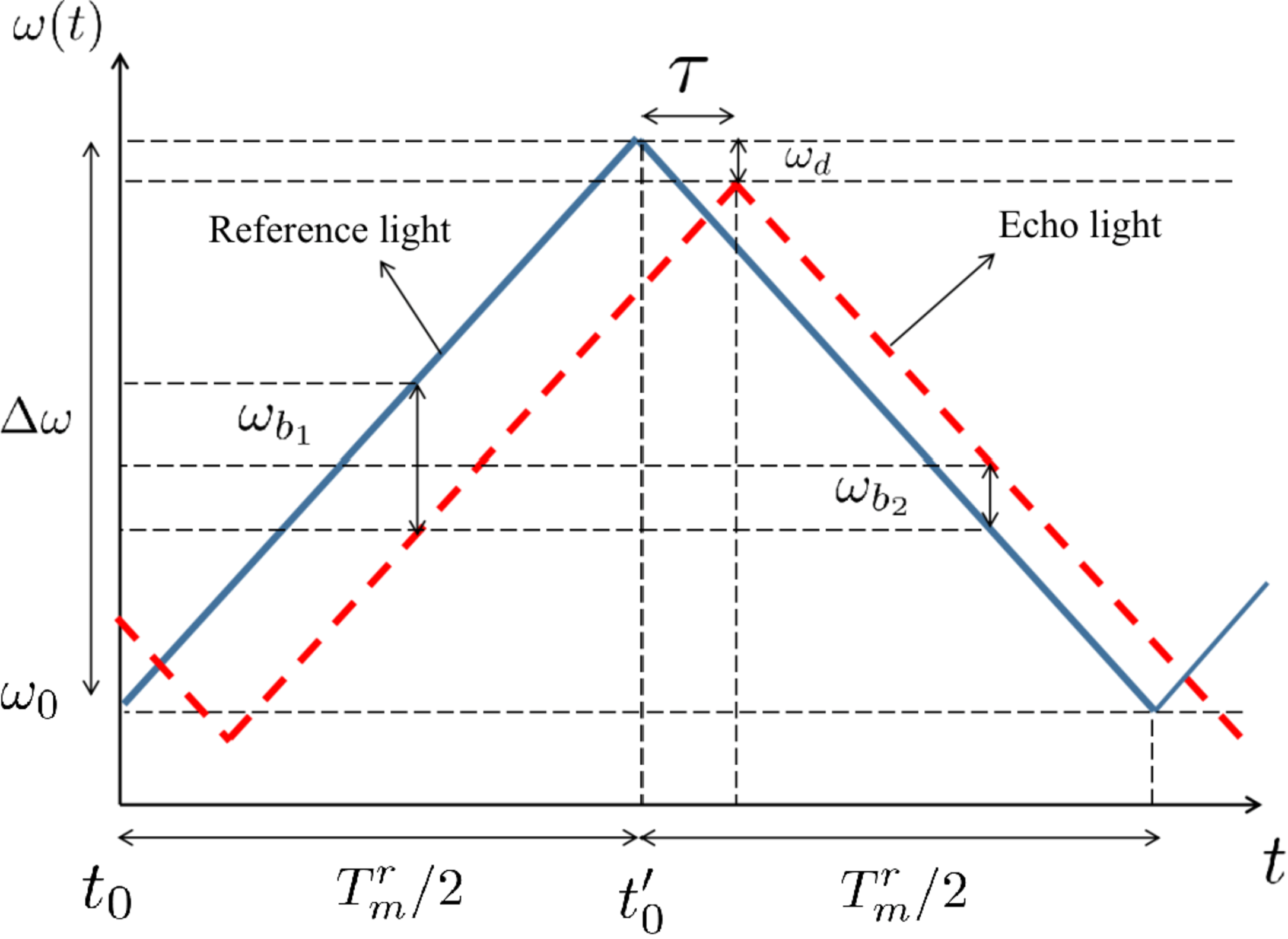}}
\caption{ The basic principle of FMCW Lidar. (a) Sawtooth FMCW Lidar target ranging, (b) Sawtooth FMCW Lidar with $\tau$ and $\omega_{d}$. (c) Triangle FMCW Lidar target ranging. (d) Triangle FMCW Lidar target ranging and velocity measurement.}
\label{fig:1}
\end{figure}

We review the basics of the classical FMCW Lidar and explain the working principle for the measurement of the target ranging and velocity with the frequency modulated continuous wave, in particular with the sawtoothed and the triangle frequency modulation as shown in Fig. \ref{fig:1}.

With the sawtoothed frequency modulation, the angular frequency of the reference light can be expressed as 
\begin{equation}\label{equ5}
\omega(t)=\frac{\Delta\omega}{T^{s}_{m}}\left(t-t_{0}\right)+\omega_{0},
\end{equation}
when $t\in[t_{0},t_{0}+T^{s}_m]$, here $t_0$ is the initial modulation time, $T^{s}_m$ is the modulation period of the sawtooth wave, $\Delta\omega$ is the modulation bandwidth, and $\omega_{0}$ is the initial angular frequency of each period. 
The angular frequency of the echo light is given by $\omega(t-\tau)\equiv\omega(t')$ with $t'\in[t_{0}+\tau,t_{0}+\tau+T^{s}_{m}]$, here $\tau=\frac{2d}{c}$ is the time delay. As shown in Fig. \ref{fig:1} (a), the echo light lags behind the reference light with a time $\tau$. In practice, $\tau\ll T^{s}_m$, and the signal is mostly obtained for $t\in [t_0+\tau, t_0+T^{s}_m]$ where the interference between the echo light and the reference light leads to a beating signal with a frequency $\omega_b$ that contains the information of $\tau$.

Specifically, $\tau$, which is just the time of flight, can be obtained from the angular frequency of the beating signal  for $t\in[t_{0}+\tau,t_{0}+T^{s}_m]$ as
 \begin{equation}
 \begin{aligned}
\tau=\frac{2d}{c}=\frac{T^{s}_{m} }{\Delta\omega}\omega_{b}=\frac{T^{s}_{m} }{\Delta\omega} [\omega(t)-\omega(t-\tau) ].
\end{aligned}
\end{equation}
Here, $2d$ is the optical path difference (OPD) between the reference light and the echo light, $c$ is the speed of light. In the free space, the target ranging, which is $d$, can then be obtained from the frequency of the beating signal.

If the target is moving, as shown in Figure \ref{fig:1}(b), the angular frequency of the beating signal, $\omega_{b}$, is not only affected by $\tau$, but also affected by the Doppler shift, denoted as $\omega_{d}$, due to the relative radial velocity $v$ between the FMCW Lidar and the target. Thus the simultaneous target ranging and velocity measurement can not be achieved with the sawtooth modulation  \cite{zhengOpticalFrequencymodulatedContinuouswave2005}. In practise, the triangle frequency modulation is often employed to achieve the simultanenous measurement of the distance and the velocity of the target.

Under the triangle frequency modulation, the frequency of the reference light, denoted as $\omega(\omega_{0},t)$, is a piecewise function that consists the rising edge and the falling edge as shown in Figure \ref{fig:1} (c) and (d). Specifically, we have 
\begin{eqnarray}\label{eq:Suq}
\omega(\omega_{0},t):=\{\begin{array}{cc}
\frac{\Delta \omega}{T_{m}^{r} / 2}\left(t-t'_{0}\right)+(\omega_{0}+\Delta \omega), & \text{for } t\in[t'_{0}-T^{r}_{m}/2,t'_{0}],\\ 
-\frac{\Delta \omega}{T_{m}^{r} / 2}\left(t-t'_{0}\right)+(\omega_{0}+\Delta \omega), & \text{for } t\in[t'_{0},t'_{0}+T^{r}_{m}/2],
\end{array}
\end{eqnarray}
Here, $T^{r}_{m}$ is the period of the triangle modulation and $t'_{0}$ is the center time of the modulation. For convenience, we choose the same modulation bandwidth, $\Delta \omega$, and the same initial angular frequency of each period, $\omega_{0}$, for both the sawtooth and the triangle modulations. And the frequency of the echo light is given by $\omega(\omega_{0}-\omega_d,t-\tau)$, which is delayed by $\tau$ and Doppler shifted by $\omega_d$. 

For a stationary target, as shown in Figure \ref{fig:1} (c), the absolute value of the beating frequency is the same at both the rising edge and the falling edge. The distance $d$ can be estimated from the beating frequencies at both edges in a similar way as the case of the sawtooth modulation. For a moving target, as shown in Fig. \ref{fig:1} (d), the frequencies of the beating signal at the rising edge and the falling edge are different. For the rising edge, the frequency of the beating signal is  
\begin{equation}
\omega_{b_{1}}=\omega_{1}(\omega_{0},t)-\omega_{1}(\omega_{0}-\omega_{d},t-\tau) =\frac{\Delta \omega}{T_{m}^{r} / 2}\tau+\omega_{d},
\end{equation}
and for the falling edge the frequency is
\begin{equation}
\omega_{b_{2}}=\omega_{2}(\omega_{0},t)-\omega_{2}(\omega_{0}-\omega_{d},t-\tau) =-\frac{\Delta \omega}{T_{m}^{r} / 2}\tau+\omega_{d}.
\end{equation}

We can then estimate both $\tau$ and $\omega_{d}$ from $\omega_{b_{1}}$ and $\omega_{b_{1}}$ as (note that the absolute value of $\omega_{b_{2}}$ is taken because $\omega_{b_{2}}<0$)
\begin{equation}\label{equ23}
\tau=\frac{T_{m}^{r} }{2\Delta \omega }\frac{\omega_{b_{1}}+|\omega_{b_{2}}|}{2} =\frac{2d}{c}
\end{equation}

\begin{equation}\label{equ24}
\omega_{d}=\frac{\omega_{b_{1}}-|\omega_{b_{2}}|}{2} = \omega_{c}\frac{2v}{c}
\end{equation}
here $\omega_{c}$ is the center frequency of the reference light, $v$ is the velocity of the target with $v<<c$ where $c$ is the speed of light. The distance and velocity of the target can then be measured simultaneously from the triangle FMCW LiDAR. 

\subsection{The measurement of the distance of the target by using FMCW LiDAR}

In addition to the target ranging and velocity measurement, FMCW LiDAR can also measure the relative distance of a target via phase estimation, similar to interferometric LiDAR based on monochromatic light.

For the sawtoothed frequency modulation, the phase of the FMCW light field  is given by
\begin{equation}
\phi(t)=\int_{t_{0}}^{t} \omega(t) d t=\omega_{0} (t-t_{0})+\frac{\Delta \omega }{2 T^{s}_{m}}(t-t_{0})^{2}
\end{equation}
for each period, here $T_m^s$ denotes the duration of a period. The phase difference between the reference light and the echo light is then
\begin{equation}
\begin{aligned}
\Delta\phi(t) &=\phi(t)-\phi(t-\tau)\\
                &=\omega_{0} \tau+\Delta \omega \tau\frac{(t-t_{0}-\tau / 2)}{T^{s}_m} \\
                &\approx (\omega_{0}-\frac{\Delta \omega}{T^{s}_{m}}t_{0}) \tau +\frac{\Delta \omega \tau}{T^{s}_m}t=(\omega_{0}-\frac{\Delta \omega}{T^{s}_{m}}t_{0}) \tau+\omega_{b}t
\end{aligned}
\end{equation}
where $\omega_{b}=\Delta \omega \tau/T^{s}_m$ and the approximation holds for $T^{s}_{m}\gg\tau$. This suggests that there exists a constant relative phase given by 
\begin{equation}
\Delta \phi_{0}=(\omega_{0}-\frac{\Delta \omega}{T^{s}_{m}}t_{0}) \tau=(\omega_{0}-\frac{\Delta \omega}{T^{s}_{m}}t_{0}) \frac{2d}{c}=2\pi \frac{2d}{\lambda_{0}},
\end{equation}
between the reference light and the echo light. The relative phase $\Delta \phi_{0}$ appears as the initial phase of the beat signal, which is only sensitive to distances within one wavelength $\lambda_{0}=2\pi c/(\omega_{0}-\frac{\Delta \omega}{T^{s}_{m}}t_{0})$. This is because the initial phase of the beat signal changes by one period for each wavelength $\lambda_{0}$ of the echo light propagation. Therefore, in the context of LiDAR ($d>>\lambda_{0}$), the initial phase of the beat signal will not contain information about the distance $d$.

The relative distance can be estimated with great resolution by watching this phase, despite the fact that the information contained in a single value of $\Delta \phi_0$ is restricted. Similar to interferometric imaging utilizing monochromatic light \cite{turbideAlloptronicSyntheticAperture2012,terrouxSyntheticApertureLidar2017}, the detailed contour of the target's surface can be acquired by measuring a collection of $\Delta \phi_0$ values from various spots. The target's velocity can also be determined using simply sawtoothed frequency modulation when the target's velocity is low enough to ignore the Doppler effect \cite{rao2017introduction}. By determining the difference of two relative phases $\Delta \phi_0$ in two consecutive modulation periods, the target's velocity in this instance is determined.

The instantaneous phase of reference light for the triangle frequency modulation is represented as a piecewise function by
\begin{equation}
\phi_{1}(t)=\int_{t}^{t'_{0}} \omega_{2}(\omega_{0},t) d t=-[(\omega_{0}+\Delta \omega)(t-t'_{0})+\frac{\Delta \omega }{ T^{r}_{m}}(t-t'_{0})^{2}]
\end{equation}
for rising edge, and 
\begin{equation}
\phi_{2}(t)=\int_{t'_{0}}^{t} \omega_{2}(\omega_{0},t) d t=(\omega_{0}+\Delta \omega)(t-t'_{0})-\frac{\Delta \omega }{ T^{r}_{m}}(t-t'_{0})^{2},
\end{equation}
for falling edge. On the other hand, for moving target, the time of flight becomes \cite{ivanovElaboratedSignalModel2020}
\begin{equation}
\tau(t)\approx\tau+\frac{\omega_{d}}{\omega_{c}} t\equiv\tau+k t=\frac{2(d+vt)}{c}
\end{equation}
where $k\equiv\omega_{d}/\omega_{c}$, if $v<<c$, i.e., $k<<1$. Hence, for the triangle frequency modulation, the instantaneous phase of the echo light in the rising edge is given by 
\begin{equation}
\begin{aligned}
\phi_{1}[t-\tau(t)]&=-[(\omega_{0}+\Delta \omega) (t-kt-\tau-t'_{0})+\frac{\Delta \omega }{ T^{r}_{m}}(t-kt-\tau-t'_{0})^{2}] \\
&=-[(\omega_{0}+\Delta \omega) (1-k)(t-\frac{\tau+t'_{0}}{1-k})+\frac{\Delta \omega }{ T^{r}_{m}/(1-k)^{2}}(t-\frac{\tau+t'_{0}}{1-k})^{2}]\\
&\approx -[(\omega_{0}+\Delta \omega) (1-k)(t-\tau-t'_{0})+\frac{\Delta \omega }{ T^{r}_{m}}(t-\tau-t'_{0})^{2}],
\end{aligned}
\end{equation}
where the approximation holds for $k<<1$. Furthermore, because of $\omega_{0} >>\Delta\omega$ in practice, the approximation
\begin{equation}
(\omega_{0}+\Delta \omega)k=\frac{(\omega_{0}+\Delta \omega)}{\omega_{c}}\omega_{d}=\frac{(\omega_{0}+\Delta \omega)}{(\omega_{0}+\Delta\omega/2)}\omega_{d}\approx \omega_{d}
\end{equation}
holds, which means
\begin{equation}
\begin{aligned}
\phi_{1}[t-\tau(t)]\approx  -[(\omega_{0}-\omega_{d}+\Delta \omega)(t-\tau-t'_{0})+\frac{\Delta \omega }{ T^{r}_{m}}(t-\tau-t'_{0})^{2}].
\end{aligned}
\end{equation}
Therefore, under the approximations $k<<1$ and $\omega_{0} >>\Delta\omega$, $\omega_{d}$ can be regard as the Doppler frequency shift in the modulation process as shown in Fig. \ref{fig:1}. The instantaneous phase of the echo light in falling edge is given by
\begin{equation}
\phi_{2}[t-\tau(t)]\approx(\omega_{0}-\omega_{d}+\Delta \omega) (t-\tau-t'_{0})-\frac{\Delta \omega }{ T^{r}_{m}}(t-\tau-t'_{0})^{2},
\end{equation}
if $k<<1$ and $\omega_{0} >>\Delta\omega$. 

Also appearing as a piecewise function is the instantaneous phase difference between the reference light and the echo light with
\begin{equation}
\begin{aligned}
\Delta\phi_{1}(t) &=\phi_{1}(t)-\phi_{1}[t-\tau(t)]\\
                         &\approx -(\omega_{0}-\omega_{d}+\Delta \omega) \tau-(\frac{\Delta \omega\tau}{T^{r}_{m}/2}+\omega_{d})(t-t'_{0}) \\
                         &=-(\omega_{0}-\omega_{d}+\Delta \omega)\tau+\omega_{b_{1}}t'_{0}-\omega_{b_{1}}t
\end{aligned}
\end{equation}
for rising edge, and 
\begin{equation}
\begin{aligned}
\Delta\phi_{2}(t) &=\phi_{2}(t)-\phi_{2}[t-\tau(t)]\\
                         &\approx (\omega_{0}-\omega_{d}+\Delta \omega) \tau+(-\frac{\Delta \omega\tau}{T^{r}_{m}/2}+\omega_{d})(t-t'_{0}) \\
                         &=(\omega_{0}-\omega_{d}+\Delta \omega)\tau-\omega_{b_{2}}t'_{0}+\omega_{b_{2}}t
\end{aligned}
\end{equation}
for falling edge, here $\omega_{b_{1}}\equiv\frac{\Delta \omega\tau}{T^{r}_{m}/2}+\omega_{d}$ and $\omega_{b_{2}}\equiv-\frac{\Delta \omega\tau}{T^{r}_{m}/2}+\omega_{d}$. 

As a result, using the triangle frequency modulation, it is also possible to calculate the target's relative distance using the beat signal's phase. The initial phase of the beat signal in each edge can be the same if $t' 0=0$, since the beat signal is a cosine function, which is an even function. However, from a detection standpoint, it is not possible to detect both beat signals at $t'_0=0$ because, even if $T^r_m>>tau$, there is still a temporal lag between the rising and falling edge. As a result, in the main text, we regard the phases of the beat signal in each edge as two separate phases, $\theta_0=\Delta \phi_1(0)$ and $\theta_1=\Delta \phi_2(0)$.

\section{The quantum FMCW LiDAR}

\subsection{Quantum description of the FMCW light fields in the discrete frequency domain}

In classical optics, an optical FMCW beam can be described as an oscillating electromagnetic field in the space-time as
\begin{equation}
E(\boldsymbol{r},t)=E(\boldsymbol{r})e^{i\phi_{0}} e^{i \int_{t_{0}}^{t} \omega(t) d t}
\end{equation}
where $\omega(t)$ is the modulating frequency, $E(\boldsymbol{r})$ is the amplitude at the space-point $\boldsymbol{r}$ and  $\phi_{0}$ is the initial phase of the light source. For simplicity, we consider the oscillation of the field at one space-point $\boldsymbol{r}$ with $\phi_{0}=0$ as
\begin{equation}\label{equ0}
E(r,t)\propto  E_{\omega_{c}}(t)=\alpha e^{i \int_{t_{0}}^{t} \omega(t) d t},
\end{equation}
where $\alpha$ and $\omega_{c}$ are the amplitude and center frequency at the space-point $\boldsymbol{r}$.

\begin{figure}
\centering
\subfigure[]{
\includegraphics[width=0.49\linewidth]{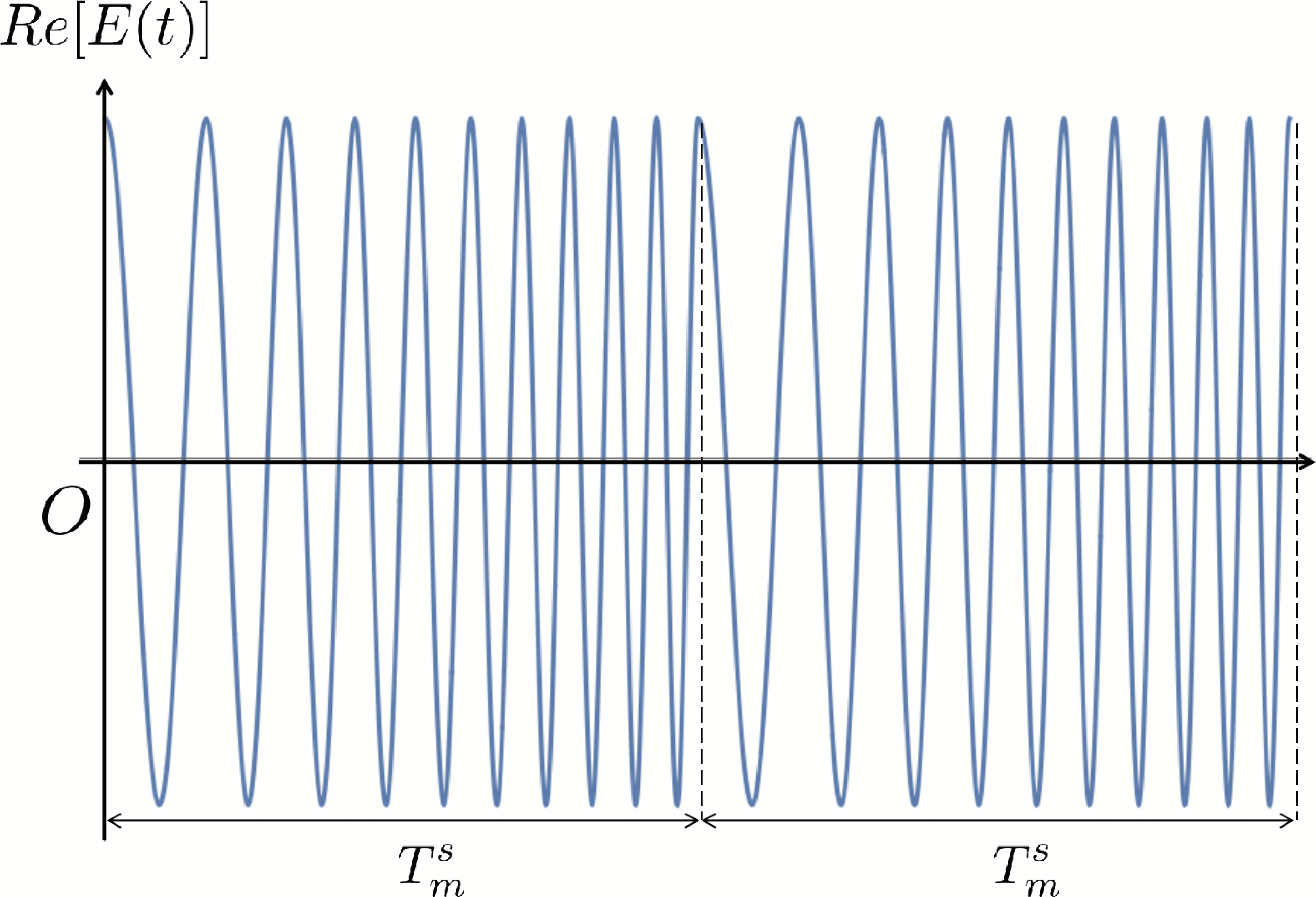}}
\subfigure[]{
\includegraphics[width=0.49\linewidth]{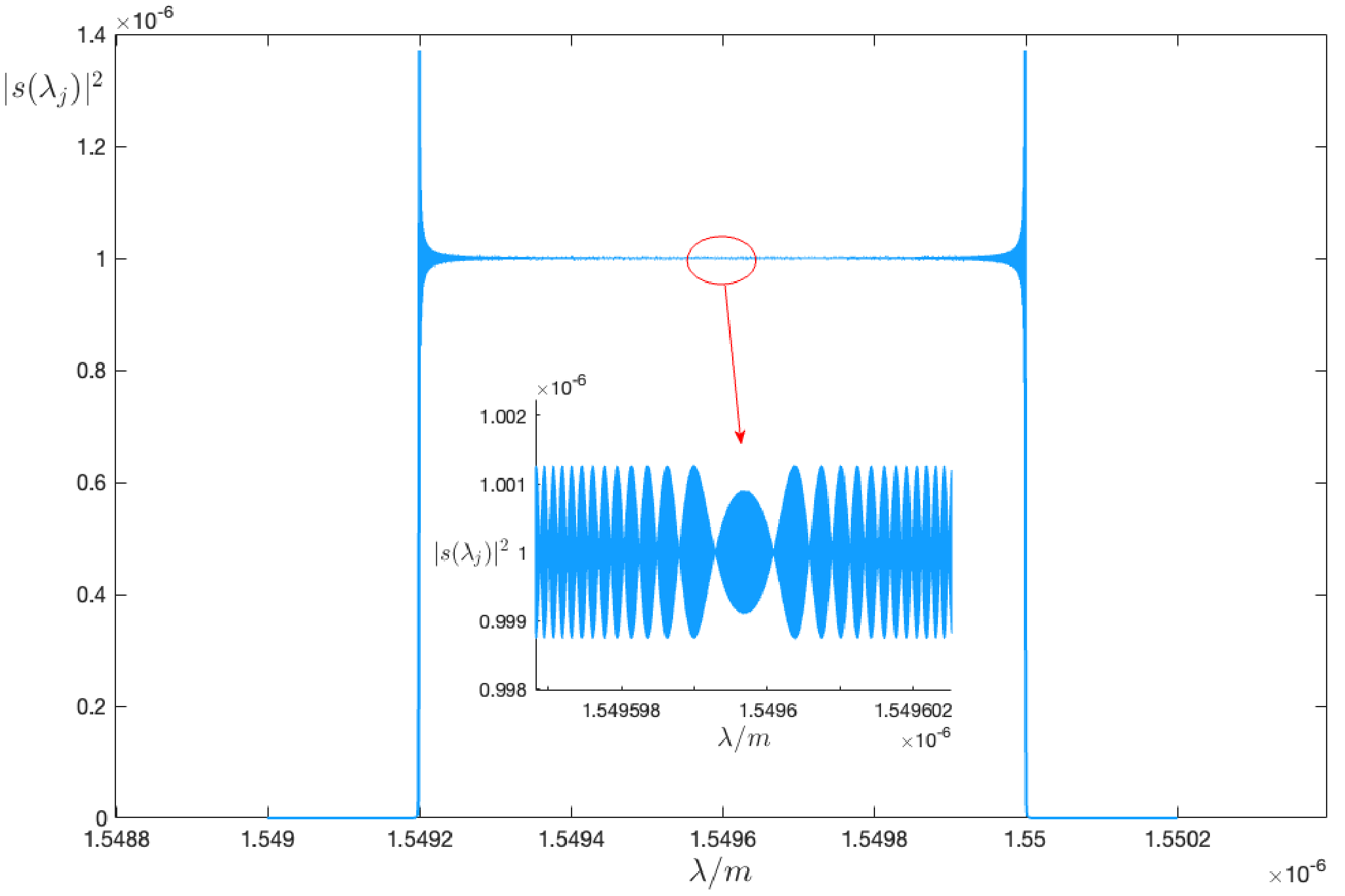}}
\caption{ (a) A classical sawtooth FMCW light field in time domain with $t_{0}=0$. (b) The power spectrum of a sawtooth FMCW light field where $\lambda_{0}=2\pi c/\omega_{0}=1550nm$, $\Delta f=\Delta \omega/2\pi=100$GHz and $T^{s}_{m}=10\mu s$. The insert shows the details of center of the spectrum }
\label{fig:0}
\end{figure}

The light field under the sawtoothed frequency modulation, as shown in Fig. \ref{fig:1} (a), can be desribed as
\begin{equation}
E_{\omega_{0}+\frac{\Delta\omega}{2}}(t)=\alpha\exp \left\{i\left[\omega_{0} t+\frac{\Delta \omega }{2 T^{s}_{m}}t^{2}\right]\right\}
\end{equation}
where we have set $t_{0}=0$, and in the first period we have $t\in [0,T^{s}_{m}]$. A sketch of the real part of $E(t)$ under the sawtoothed frequency modulation can be seen in Fig. \ref{fig:0} (a).

Since the light field is being modulated periodically, it has a discrete frequency spectrum which can be obtained from the Fourier transformation as
\begin{equation}
s_{\omega_{c}}(\omega_{j})=\frac{1}{T_{m}} \int_{t_{0}}^{t_{0}+T_{m}} E(t) e^{-i  \omega_{j} t} d t,
\end{equation}
here $T_{m}$ is the period of the modulation, $\omega_{c}$ is the center frequency, and $\omega_{j}=2\pi j/T_{m}$ is the discrete frequency under the modulation. As shown in Fig. \ref{fig:0} (b), the complex spectrum, $s_{\omega_{c}}(\omega_{j})$, is rather complicated in general.

Under the approximation of slowly varying envelope, $\omega_{c}>>\Delta\omega$ \cite{tsangQuantumTheoryOptical2008}, the classical FMCW light field, $E_{\omega_{c}}(t)$, can be described with a multi-mode coherent state as
\begin{equation}
\begin{aligned}
\mathop{\otimes}_{j}\left|\alpha s_{\omega_{c}}^{*}\left(\omega_{j}\right)\right\rangle &=\prod_{j} \exp \left[\alpha s_{\omega_{c}}^{*}\left(\omega_{j}\right) a^{+}\left(\omega_{j}\right)-\alpha^{*} s_{\omega_{c}}\left(\omega_{j}\right) a\left(\omega_{j}\right)\right]|0\rangle \\
&=\exp \left\{\sum_{j}\left[\alpha s_{\omega_{c}}^{*}\left(\omega_{j}\right) a^{+}\left(\omega_{j}\right)-\alpha^{*} s_{\omega_{c}}\left(\omega_{j}\right) a\left(\omega_{j}\right)\right]\right\}|0\rangle,
\end{aligned}
\end{equation}
where $\hat{a}(\omega_{j})$ and $\hat{a}^{+}(\omega_{j})$ are the annihilation and generation operators of the light field of the frequency $\omega_{j}$, which satisfy the commutation relation, $\left[a\left(\omega_{j}\right), a^{+}\left(\omega_{j^{\prime}}\right)\right]=\delta_{j, j^{\prime}}$. Here, we assume that the complex spectrum $s(\omega_{j})$ has been normalized, i.e., $\sum_{j}|s_{\omega_{c}}^{*}\left(\omega_{j}\right)|^{2}=1$, and $\alpha$ is the amplitude of the single-mode coherent state before the modulation. The modulation process can be described by the transformation of the operators as
\begin{equation}
\begin{aligned}
a^{+}(\omega) &\rightarrow \sum_{j} s_{\omega_{c}}^{*}\left(\omega_{j}\right) a^{+}\left(\omega_{j}\right), & a(\omega) \rightarrow \sum_{j} s_{\omega_{c}}\left(\omega_{j}\right) a\left(\omega_{j}\right)
\end{aligned}
\end{equation}
here $\omega$ is an angular frequency before the modulation.

A single-photon state of the FMCW field can then be described as
\begin{equation}\label{equ3}
|\psi_{1}\rangle=\sum_{j} s_{\omega_{c}}^{*}\left(\omega_{j}\right) a^{+}\left(\omega_{j}\right)|0\rangle,
\end{equation}
which is in the superposition of the states with the spectrum $s(\omega_{j})$, similar to the single-photon state with a Gaussian spectrum.

Different from the single photon state, it is necessary to consider the correlation of the entangled biphoton state under the frequency modulation. The entangled biphoton state with the anti-frequency-correlation is 
\begin{equation}
|\Psi_{2}\rangle=\int d \omega G(\omega) a_{1}^{+}\left(\omega_{c}+\omega\right) a_{2}^{+}\left(\omega_{c}-\omega\right)|0\rangle,
\end{equation}
where $G(\omega)$ is a Gaussian spectrum with a standard deviation $\sigma$, $\omega_{c}$ is the center frequency. Under the frequency modulation the state becomes
\begin{equation}\label{equ4}
\begin{array}{r}
\left|\psi_{2}\right\rangle=\int d \omega G(\omega) \sum_{m, n} s_{\omega_{c}+\omega}^{*}\left(\omega_{m}\right) s_{\omega_{c}-\omega}^{*}\left(\omega_{n}\right)
a_{1}^{+}\left(\omega_{m}\right) a_{2}^{+}\left(\omega_{n}\right)|0\rangle.
\end{array}
\end{equation}

\subsection{Quantum description of the FMCW light fields in the discrete time domain}

For the FMCW light field, an annihilation operator $\hat{A}(t)$ at a certain time $t$ can be described by
\begin{equation}
\hat{A}(t) = \sum_{j} \hat{a}(\omega_{j}) e^{-i \omega_{j} t},
\end{equation}
and $\hat{A}(t+T_{m})=\hat{A}(t)$. 
Due to the periodicity of $\hat{A}(t)$, without loss of generality below we will restrict $t\in [t_{0},t_{0}+T_{m}]$. Since $\left[a\left(\omega_{j}\right), a^{+}\left(\omega_{j^{\prime}}\right)\right]=\delta_{j, j^{\prime}}$, we have
\begin{equation}
\left[\hat{A}(t), \hat{A}^{+}\left(t^{\prime}\right)\right]=\sum_{j} \exp \left[i \omega_{j}\left(t^{\prime}-t\right)\right],
\end{equation}
here $t,t'\in [t_{0},t_{0}+T_{m}]$. We note that $\hat{A}(t)$ and $\hat{A}^{+}(t’)$ do not always commute when $t\neq t’$, which puts an obstacle for the modeling of the quantum FMCW light field in the time domain.

In order to overcome this obstacle, we opt to use a finite bandwidth and a discrete time approximation. This involves dividing the period of the modulation, $T_m$, into $j_B$ intervals with
\begin{equation}
t_{p}=\frac{T_{m}}{j_{B}} p, \quad p\in\left\{p_{0},p_{0}+1,p_{0}+2, \ldots p_{0}+j_{B}\right\},
\end{equation}
while the bandwidth of the frequency is limited to $\omega_{j_{B}}=2\pi j_{B}/T_{m}$, i.e., the discrete frequencies are
\begin{equation}
\omega_{j}=\frac{2\pi j}{T_{m}}, \quad j\in\left\{j_{c}-\frac{j_{B}}{2}, \ldots ,j_{c}, \ldots ,j_{c}+\frac{j_{B}}{2}\right\},
\end{equation}
here $\omega_{j_{c}}=\frac{2\pi j_c}{T_{m}}$ represents a frequency that is closest to the central frequency of the FMCW light field. The commutation relations of $\{\hat{A}(t_{p})\}$ are then 
\begin{equation}\label{equ2}
\left[\hat{A}(t_{p}), \hat{A}^{+}\left(t_{p'}\right)\right]=\left(j_{B}+1\right) \delta_{p, p'},
\end{equation}
where $j_{B}+1$ is the dimension of the discrete time as well as the discrete frequencies within the bandwidth. With the approximation $\hat{A}(t_{p})$ and $\hat{A}^{+}(t_{p’})$ always commute for $p\neq p’$. Furthermore, the commutation relation Eq. (\ref{equ2}) can be normalized by letting
\begin{equation}\label{equ8}
\hat{a}(t_{p}) \equiv \frac{\hat{A}(t_{p})}{\sqrt{j_{B}+1}}=\frac{1}{\sqrt{j_{B}+1}} \sum_{j=j_{c}-\frac{j_{B}}{2}}^{j_{c}+\frac{j_{B}}{2}} \hat{a}\left(\omega_{j}\right) \exp \left(-i \omega_{j} t_{p}\right),
\end{equation}
which satisfies the canonical commutation relation  $\left[\hat{a}(t_{p}), \hat{a}^{+}\left(t_{p’}\right)\right]=\delta_{p, p^{\prime}}$.
We can also obtain the operators in the frequency domain as
\begin{equation}
\hat{a}\left(\omega_{n}\right)=\frac{1}{\sqrt{j_{B}+1}} \sum^{p_{0}+j_{B}}_{p=p_{0}} \hat{a}\left(t_{p}\right) e^{i \omega_{n} t_{p}},
\end{equation}
which forms a pair in the discrete Fourier transformation with $\hat{a}(t_p)$. In this case the light intensity operator at $t_{p}$ is
\begin{equation}
\hat{E}^{(-)}(t_{p})\hat{E}^{(+)}(t_{p}) \propto \hat{A}^{+}(t_{p})\hat{A}(t_{p})= (j_{B}+1)a^{+}(t_{p})a(t_{p}).
\end{equation}

In analogy to the Fock state in the discrete frequency mode, we can define a Fock state in the discrete time mode \cite{tsangQuantumTheoryOptical2008} as
\begin{equation}
\left|n_{t_{p}}\right\rangle=\frac{1}{\sqrt{n_{t_{p}} !}}\left[\hat{a}^{+}(t_{p})\right]^{n_{t_{p}}}|0\rangle,
\end{equation}
here $n_{t_{p}}$ is the photon number at the discrete time $t_{p}$. As shown in \cite{tsangQuantumTheoryOptical2008}, the band-limited Hilbert space can be spanned by this discrete time mode as
\begin{equation}
I=\otimes_{t_{p}} \sum_{n_{t_{p}}}\left|n_{t_{p}}\right\rangle\left\langle n_{t_{p}}\right|=\sum_{\mathbf{n}}\left|\mathbf{n}\right\rangle\left\langle \mathbf{n}\right|,
\end{equation}
here $|\mathbf{n}\rangle=|n_{t_{p_{0}}},\ldots,n_{t_{p_{0}}+T_{m}}\rangle$ is a photon number state in a period.

The single FMCW photon state, as shown in Eq.(\ref{equ3}), can be rewritten in the discrete time domain as
\begin{equation}\label{equ6}
\begin{aligned}
& |\psi_{1}\rangle=\sum_{j} s_{\omega_{c}}^{*}\left(\omega_{j}\right) a^{+}\left(\omega_{j}\right)|0\rangle \\
&\approx \frac{1}{\sqrt{j_{B}+1}} \sum_{j=j_{c}-\frac{j_{B}}{2}}^{n_{c}+\frac{j_{B}}{2}} \sum_{p=p_{0}}^{p_{0}+j_{B}} s_{\omega_{c}}^{*}\left(\omega_{j}\right) a^{+}\left(t_{p}\right) e^{-i \omega_{j} t_{p}}|0\rangle \\
& \approx  \frac{1}{\sqrt{j_{B}+1}} \sum_{p=p_{0}}^{p_{0}+j_{B}} E_{\omega_{c}}^{*}\left(t_{p}\right) a^{+}\left(t_{p}\right)|0\rangle \equiv |\psi’_{1}\rangle ,
\end{aligned}
\end{equation}
here $E_{\omega_{c}}\left(t_{p}\right)$ is the FMCW light field given in Eq. (\ref{equ0}) with $\alpha=1$, and we assume $j_{B}$ is sufficiently large to include most frequency band of the FMCW light field. When $j_{B}\rightarrow\infty$, the single photon state becomes
\begin{equation}
|\psi^{\prime\prime}_{1}\rangle=\frac{1}{\sqrt{T_{m}}} \int^{t_{0}+T_{m}}_{t_0} E_{\omega_{c}}^{*}(t) a^{+}(t)|0\rangle d t
\end{equation}
which is expected without the approximation of a finite bandwidth.

The entangled biphoton FMCW state, as in Eq.(\ref{equ4}), can also be rewritten in the discrete time domain as
\begin{equation}\label{equ7}
\begin{aligned}
&\left|\psi_{2}\right\rangle=\int d \omega G(\omega) \sum_{m, n} s_{\omega_{c}+\omega}^{*}\left(\omega_{m}\right) s_{\omega_{c}-\omega}^{*}\left(\omega_{n}\right) a_{1}^{+}\left(\omega_{m}\right) a_{2}^{+}\left(\omega_{n}\right)|0\rangle \\
&\approx \frac{1}{j_{B}+1} \int d \omega G(\omega) \sum^{p_{0}+j_{B}}_{p, q=p_{0}} E_{\omega_{c}+\omega}^{*}\left(t_{p}\right) E_{\omega_{c}-\omega}^{*}\left(t_{q}\right)  a_{1}^{+}\left(t_{p}\right) a_{2}^{+}\left(t_{q}\right)|0\rangle \\
&=\frac{1}{j_{B}+1}  \sum^{p_{0}+j_{B}}_{p, q=p_{0}} \int d \omega G(\omega) e^{-i \omega\left(t_{p}-t_{q}\right)} E_{\omega_{c}}^{*}\left(t_{p}\right) E_{\omega_{c}}^{*}\left(t_{q}\right)a_{1}^{+}\left(t_{p}\right) a_{2}^{+}\left(t_{q}\right)|0\rangle \\
&=\frac{1}{j_{B}+1}  \sum^{p_{0}+j_{B}}_{p, q=p_{0}} \tilde{G}\left(t_{p}-t_{q}\right) E_{\omega_{c}}^{*}\left(t_{p}\right) E_{\omega_{c}}^{*}\left(t_{q}\right) a_{1}^{+}\left(t_{p}\right) a_{2}^{+}\left(t_{q}\right)|0\rangle\\
&\equiv |\psi’_{2}\rangle
\end{aligned}
\end{equation}
here $\tilde{G}\left(t\right)$ is the Fourier transformation of the Gaussian spectrum, $G(\omega)$. If one abandons the assumption of the finite bandwidth with $j_{B}\rightarrow\infty$, the state is
\begin{equation}
\begin{aligned}
|\psi^{\prime\prime}_{2}\rangle=\frac{1}{T_{m}} \int^{t_{0}+T_{m}}_{t_0}\int^{t_{0}+T_{m}}_{t_0} d t_{1} d t_{2} \tilde{G}\left(t_{1}-t_{2}\right)& E_{\omega_{c}}^{*}\left(t_{1}\right) E_{\omega_{c}}^{*}\left(t_{2}\right) a_{1}^{+}(t_{1}) a_{2}^{+}(t_{2})|0\rangle.
\end{aligned}
\end{equation}
Furthermore,  when $\tilde{G}\left(t_{1}-t_{2}\right)=\delta(t_{1}-t_{2})$, i.e., the variance of the $G(\omega)$ becomes infinity, $\sigma\rightarrow\infty$, the state is
\begin{equation}
\begin{aligned}
|\psi^{\prime\prime\prime}_{2}\rangle=\frac{1}{\sqrt{T_m}} \int_{t_{0}}^{t_{0}+T_{m}} E_{\omega_{c}}^{* 2}\left(t\right) a_{1}^{+}\left(t\right)a_{2}^{+}\left(t\right)|0\rangle,
\end{aligned}
\end{equation}
in this case the two photons is not only entangled but also synchronized.

\subsection{Propagation of the quantum FMCW Light field}

\begin{figure}[b]
\centering
\includegraphics[width=0.6\linewidth]{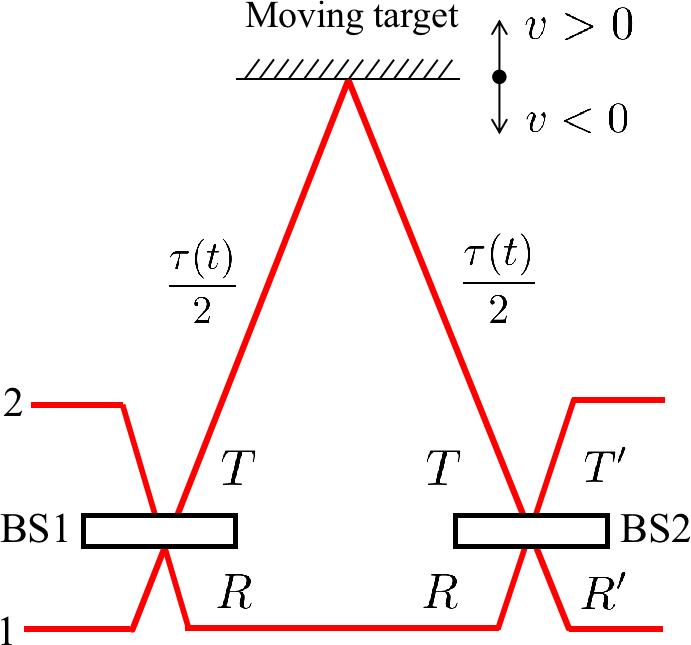}
\caption{The sketch of target ranging and velocity measurement by using MZ interferometer. The BS1 and BS2 are the beam splitter.}
\label{fig:2}
\end{figure}

The propagation of the FMCW field in the target ranging and velocity measurement is shown in Fig. \ref{fig:2}. The input quantum FMCW light field is split by BS1 into the reference and the signal. The signal is then sent to interrogate a moving target, reflected as the echo light. The echo light then interferes with the reference light through BS2. The scheme is similar to the Mach-Zehnder (MZ) interferometer, but has two differences. First, the target is always placed in the middle of one arm to create a specular-like reflection that simulates the round-trip propagation of the signal. Second, the evolution of the light field is described by the frequency-independent time of light ($\tau(t)\approx\tau+(\omega_{d}/\omega_{c})t=2(d+vt)/c$ (if $|v|<<c$)) \cite{ivanovElaboratedSignalModel2020} instead of the frequency-dependent phase differences typically used in the MZ interferometer. Note that we define the target velocity $v>0$ ($v<0$) when the target is moving away (toward) from the photon detectors that creates a red (blue) shift.

BS1 and BS2 are 50:50 beam splitters with the input-output relation \cite{cable2007efficient} 
\begin{equation}
\begin{aligned}
&a_{1}^{+}\left(\omega\right) \rightarrow \frac{1}{\sqrt{2}}\left[a_{R}^{+}\left(\omega\right)+a_{T}^{+}\left(\omega\right)\right], \\
&a_{2}^{+}\left(\omega\right) \rightarrow \frac{1}{\sqrt{2}}\left[a_{R}^{+}\left(\omega\right)-a_{T}^{+}\left(\omega\right)\right].
\end{aligned}
\end{equation}
Due to the linearity of the discrete Fourier transform, as shown in Eq. (\ref{equ8}), the FMCW light field in the time domain also has a similar relation:
\begin{equation}\label{equ9}
\begin{aligned}
&a_{1}^{+}(t_{p}) \rightarrow \frac{1}{\sqrt{2}}\left[a_{R}^{+}(t_{p})+a_{T}^{+}(t_{p})\right], \\
&a_{2}^{+}(t_{p}) \rightarrow \frac{1}{\sqrt{2}}\left[a_{R}^{+}(t_{p})-a_{T}^{+}(t_{p})\right].
\end{aligned}
\end{equation}
The evolution of $a_{T}^{+}(t)$ in the free space can be similarly obtained from the free evolution of each $a_{T}^{+}(\omega_{j})$, which is given by
\begin{equation}
a_{T}^{+}[t+\tau(t)]=\frac{1}{\sqrt{j_{B}+1}} \sum_{j=n_{c}-\frac{j_{B}}{2}}^{n_{c}+\frac{j_{B}}{2}} a_{T}^{+}\left(\omega_{j}\right) \exp \left\{i \omega_{j}[t+\tau(t)]\right\}.
\end{equation}

In the case of using separable photons, a single FMCW state, $|\psi_{1}\rangle$, is input into the port 1 of BS1 and a vacuum state enters port 2. Before BS2 the state $|\psi’_{1}\rangle$ becomes
\begin{equation}
\begin{aligned}
|\psi’_{1}\rangle&=\frac{1}{\sqrt{2\left(j_{B}+1\right)}} \sum^{p_{0}+j_{B}}_{p=p_{0}}\left \{E_{\omega_{c}}^{*}\left(t_{p}\right) a_{R}^{+}\left(t_{p}\right)+ E_{\omega_{c}}^{*}\left[t_{p}-\tau\left(t_{p}\right)\right] a_{T}^{+}\left(t_{p}\right)\right\}|0\rangle \\
& \approx \frac{1}{\sqrt{2\left(j_{B}+1\right)}}  \sum^{p_{0}+j_{B}}_{p=p_{0}}\left \{E_{\omega_{c}}^{*}\left(t_{p}\right) a_{R}^{+}\left(t_{p}\right)+ E_{\omega_{c}-\omega_{d}}^{*}\left(t_{p}-\tau\right) a_{T}^{+}\left(t_{p}\right)\right\}|0\rangle,
\end{aligned}
\end{equation}
here the approximation holds for $\omega_{c}\gg\Delta \omega$ and $c\gg v$, as presented in Sec. II.

In the case of using entangled FMCW biphoton state, $|\psi_{2}\rangle$, a photon is inputed into the port 1 and port 2 simultaneously. Before BS2 the state $|\psi'_{2}\rangle$ becomes
\begin{equation}\label{equ10}
\begin{aligned}
&|\psi'_{2}\rangle\approx\frac{1}{2\left(j_{B}+1\right)}  \sum^{p_{0}+j_{B}}_{p,q=p_{0}}\tilde{G}\left(t_{p}-t_{q}\right)\left\{E_{\omega_{c}}^{*}\left(t_{p}\right) E_{\omega_{c}}^{*}\left(t_{q}\right) a_{R}^{+}\left(t_{p}\right) a_{R}^{+}\left(t_{q}\right)-\right.\\
&\enspace\enspace\enspace\enspace\enspace\enspace\enspace\enspace\enspace\enspace\enspace\enspace\enspace\enspace\enspace\enspace\enspace\enspace\enspace\enspace\enspace\left.E_{\omega_{c}-\omega_{d}}^{*}\left(t_{p}-\tau\right) E_{\omega_{c}-\omega_{d}}^{*}\left(t_{q}-\tau\right) a_{T}^{+}\left(t_{p}\right) a_{T}^{+}\left(t_{q}\right)\right\}|0\rangle.
\end{aligned}
\end{equation}

Without the approximation of limited bandwidth, the summation can be replaced by integral and the states $|\psi'_{1}\rangle$  can be written as 
\begin{equation}
\begin{aligned}
\left|\psi_{1}^{\prime \prime}\right\rangle= \frac{1}{\sqrt{2 T_{m}}} \int^{t_{0}+T_{m}}_{t_0} d t&\left\{E_{\omega_{c}}^{*}(t)\left|1_{t}, 0\right\rangle+E_{\omega_{c}-\omega_{d}}^{*}(t-\tau) \left|0,1_{t}\right\rangle\right\},
\end{aligned}
\end{equation}
and with $\tilde{G}\left(t_{p}-t_{q}\right)=\delta(\left(t_{p}-t_{q}\right)$, $|\psi'_{2}\rangle$ can be written as
\begin{equation}
\begin{aligned}
\left|\psi_{2}^{\prime \prime }(\boldsymbol{O})\right\rangle=\frac{1}{\sqrt{2 T_{m}}} \int^{t_{0}+T_{m}}_{t_0} d t&\left\{E_{\omega_{c}}^{* 2}(t)\left|2_{t}, 0\right\rangle-E_{\omega_{c}-\omega_{d}}^{* 2}(t-\tau) \left|0,2_{t}\right\rangle\right\}.
\end{aligned}
\end{equation}
And it is straightforward to extend the biphoton FMCW NOON state to the $n$-photon FMCW NOON state as
\begin{equation}\label{equ11}
\begin{aligned}
\left|\psi''_{n}\right\rangle=\frac{1}{\sqrt{ 2T_{m}}} \int^{t_{0}+T_{m}}_{t_{0}} d t&\left\{E_{\omega_{c}}^{* n}(t)\left|n_{t}, 0\right\rangle-E_{\omega_{c}-\omega_{d}}^{* n}(t-\tau) \left|0,n_{t}\right\rangle\right\},
\end{aligned}
\end{equation}
The complexity of generating the FMCW NOON state, however, grows exponentially with $n$.

\subsection{Detection of quantum FMCW Light field for target ranging and velocity measurement\label{IID}} 

\begin{figure}[b]
\centering
\includegraphics[width=0.7\linewidth]{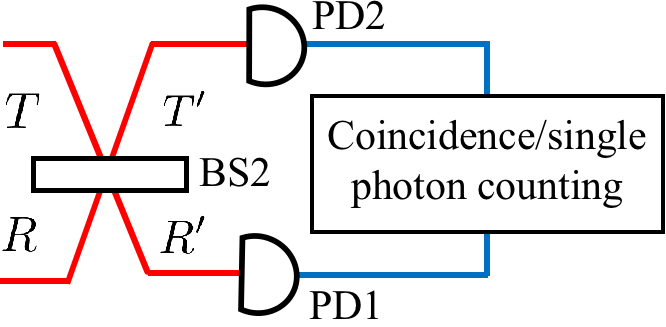}
\caption{The detection process of quantum FMCW Light field for target ranging and velocity measurement. The PD1 and PD2 are the photon detectors. For separable photon state, $|\psi_{1}\rangle$, the single photon counting is used for the detection. For entangled biphoton FMCW state, $|\psi_{2}\rangle$, the coincidence photon counting is used. }
\label{fig:4}
\end{figure}

The detection scheme consists of a beam splitter and two photon detectors for coincidence counting, as shown in Fig. \ref{fig:4}. The absorption of the photons at the photon detectors, under the limited bandwidth approximation, can be represented by the positive frequency light field operators
\begin{equation}
\begin{aligned}
\hat{E}_{R^{\prime}}^{(+)}\left(t_{1}\right) & \propto \sqrt{j_{B}+1} a_{R^{\prime}}\left(t_{1}\right) \rightarrow \sqrt{j_{B}+1} \frac{1}{\sqrt{2}}\left[a_{R}\left(t_{1}\right)+a_{T}\left(t_{1}\right)\right]
\end{aligned}
\end{equation}
and
\begin{equation}
\begin{aligned}
\hat{E}_{T^{\prime}}^{(+)}\left(t_{2}\right) & \propto \sqrt{j_{B}+1} a_{T^{\prime}}\left(t_{2}\right) \rightarrow \sqrt{j_{B}+1} \frac{1}{\sqrt{2}}\left[a_{R}\left(t_{2}\right)-a_{T}\left(t_{2}\right)\right]
\end{aligned}
\end{equation}
Here, $t_{1}$ $(t_{2})$ is the time of the photon hitting the photon detector PD1 (PD2). 

For separable photon state,  we use the photon counting at each output ports to keep track of the variation of the detection probability. Then, for the single photon state, $|\psi'_{1}(\vec{x})\rangle$ with $\vec{x}$ denotes the unknown parameters to be estimated, in our case $\vec{x}=(d,v,\theta_0,\theta_1)$. The detecting probability of PD1 at time $t$ can be obtained from the first order quantum correlation function as 
\begin{equation}
p_{1}(1|t,\vec{x})\equiv\langle\psi'_{1}(\vec{x})|\hat{E}_{R^{\prime}}^{(-)}\left(t\right) \hat{E}_{R^{\prime}}^{(+)}\left(t\right)| \psi'_{1}(\vec{x})\rangle,
\end{equation}
and the detecting probability of PD2 is
\begin{equation}
\begin{aligned}
p_{1}(0|t,\vec{x})&\equiv\langle\psi'_{1}(\boldsymbol{O})|\hat{E}_{T^{\prime}}^{(-)}\left(t\right) \hat{E}_{T^{\prime}}^{(+)}\left(t\right)| \psi'_{1}(\vec{x})\rangle =1-p_{1}(1|t,\vec{x}).
\end{aligned}
\end{equation}

For entangled biphoton FMCW state, we use the coincidence counting of the two photon counters to keep track of the variation of the joint detecting probability. For the entangled biphoton state, $|\psi'_{2}(\vec{x})\rangle$, the joint probability of the coincidence counting can be obtained from the second order quantum correlation function as
\begin{equation}
p_{2}\left(1 \mid t, \vec{x}\right)=\langle\psi'_{2}( \vec{x})|\hat{E}_{R^{\prime}}^{(-)}\left(t_{1}\right) \hat{E}_{T^{\prime}}^{(-)}\left(t_{2}\right) \hat{E}_{T^{\prime}}^{(+)}\left(t_{2}\right) \hat{E}_{R^{\prime}}^{(+)}\left(t_{1}\right)| \psi'_{2}( \vec{x})\rangle
\end{equation}
with a time delay $\tau'=t_{1}-t_{2}$. Given that the time delay is normally distributed as in Eq. (\ref{equ10}), the probability of the coincidence counting becomes 
\begin{equation}
\begin{aligned}
p_{2}\left(1 \mid t,  \vec{x}\right)=\int d \tau^{\prime}&\langle\psi'_{2}( \vec{x})|\hat{E}_{R^{\prime}}^{(-)}\left(t\right) \hat{E}_{T^{\prime}}^{(-)}\left(t-\tau^{\prime}\right) \hat{E}_{T^{\prime}}^{(+)}\left(t-\tau^{\prime}\right) \hat{E}_{R^{\prime}}^{(+)}\left(t\right)| \psi'_{2}( \vec{x})\rangle,
\end{aligned}
\end{equation}
where one of the photon is detected by PD1 at time $t_1=t$ and the other photon is detected by PD2 with arbitrary time delay. Thus the probability of both photons detected by the same detector (PD1 or PD2) is
\begin{equation}
p_{2}\left(0 \mid t,  \vec{x}\right)=1-p_{2}\left(1 \mid t,  \vec{x}\right).
\end{equation}

\begin{figure}[h]
\centering
\includegraphics[width=0.8\linewidth]{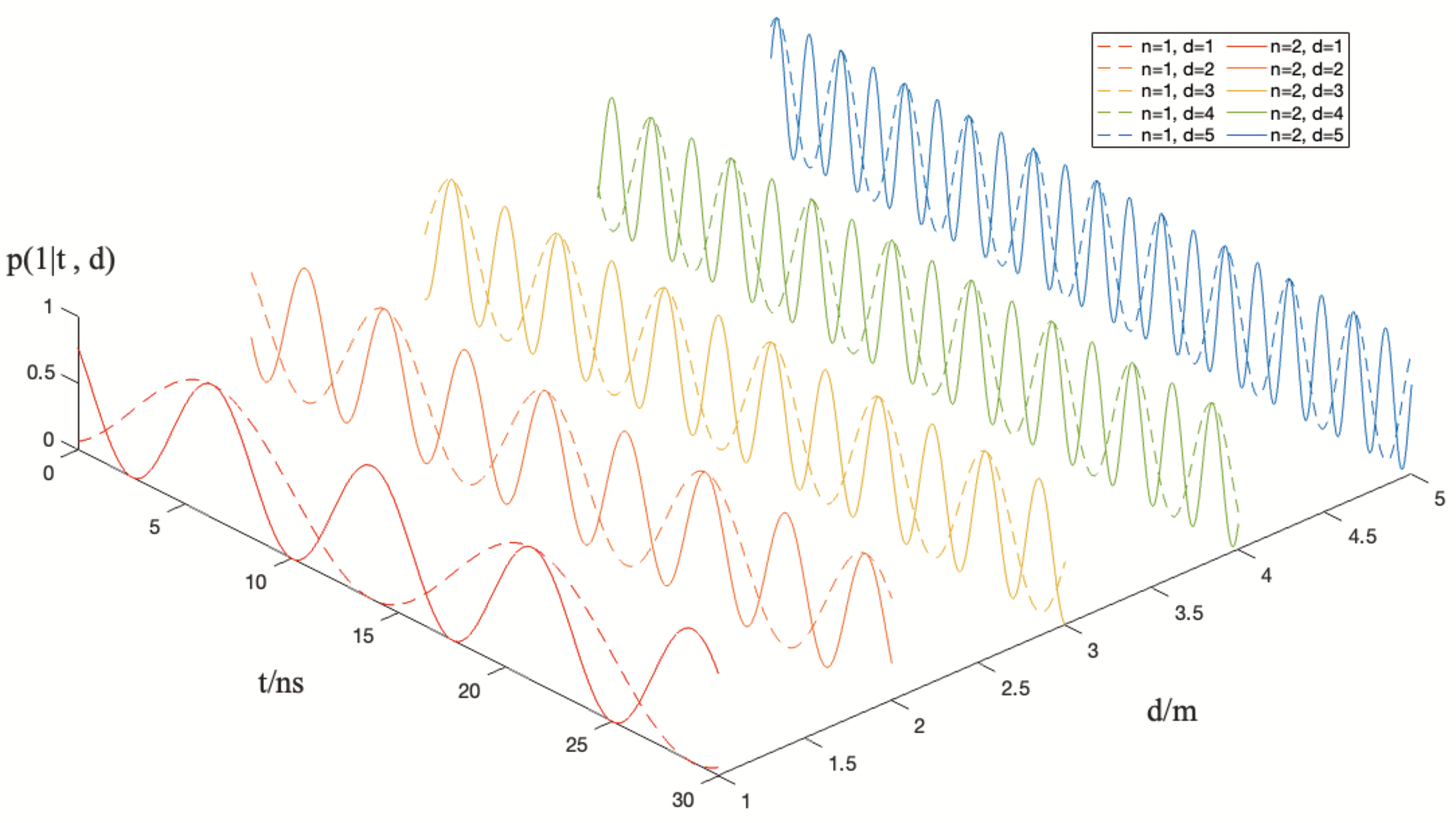}
\caption{A simulation of the quantum beating signal $p_{n}\left(1 \mid t, d\right)$ for $n=1$ (dashed line) and $n=2$ (solid line) where $t_{d}=0$, $\lambda_{0}=2\pi c/\omega_{0}=1550nm$, $\Delta f=\Delta \omega/2\pi=100$GHz and $T^{s}_{m}=10\mu s$. Note that, the relative phase $\theta$ is fixed if the absolute distance $d$ is certain.}
\label{fig:3}
\end{figure}

For $v=0$, the sawtooth modulation can be used, which yields a probability of detection for each modulation period given by
\begin{equation}\label{equ26}
p_{n}\left(1 \mid t, d, \theta\right) \approx \frac{1}{2}\left[1+\cos \left(n \frac{\Delta \omega}{T^{s}_{m}} \frac{2 d}{c} t+n \theta\right)\right],
\end{equation}
where $n=1,2$ corresponds to the single-photon and biphoton entanglement states, respectively. The approximation holds if $T^{s}_{m}\gg\tau$ and $\sigma\rightarrow \infty$. Here, $\theta=(\omega_{0}-\frac{\Delta \omega}{T_{m}}t_{d}) \frac{2d}{c}$ with an initial detection time $t_{d}$. The probability of detection oscillates with time $t$, creating a quantum beat signal, as shown in Fig. \ref{fig:3}. This oscillation frequency increases with the target's distance $d$. Thus, the target's distance $d$ can be determined from the frequency of the beat signal. The beat signal has the same form as that provided by an FMCW laser when $n=1$. In the biphoton scenario where $n=2$, the frequency and initial phase of the beat signal are double that of the single-photon scenario ($n=1$).

For $v\neq 0$, the triangular modulation can be used, which yields a piecewise probability of detection for $n=1,2$ as
\begin{equation}\label{equ12}
\begin{aligned}
&p_{n}(1 \mid d, v, \theta_{0}, t) \approx  \frac{1}{2}\left\{1+\cos \left[n\left(\frac{\Delta \omega}{T^{r}_{m}/2} \frac{2 d}{c}+\omega{c} \frac{2 v}{c}\right) t+n \theta_{0} \right]\right\},
\end{aligned}
\end{equation}
where
\begin{equation}
\begin{aligned}
\theta_{0}= (\omega_{0}-\omega_{c} \frac{2 v}{c}+\Delta \omega) \frac{2 d}{c}-\left(\frac{\Delta \omega}{T^{r}_{m}/2} \frac{2 d}{c}+\omega{c} \frac{2 v}{c}\right)t_{d_{0}}
\end{aligned}
\end{equation}
for rising edge, and
\begin{equation}\label{equ13}
\begin{aligned}
&p_{n}(1 \mid d, v, \theta_{1}, t) \approx \frac{1}{2}\left\{1+\cos \left[n\left(-\frac{\Delta \omega}{T^{r}_{m}/2} \frac{2 d}{c}+\omega{c} \frac{2 v}{c}\right) t+n \theta_{1}\right]\right\},
\end{aligned}
\end{equation}
where
\begin{equation}
\begin{aligned}
\theta_{1}= (\omega_{0}-\omega_{c} \frac{2 v}{c}+\Delta \omega) \frac{2 d}{c}-\left(-\frac{\Delta \omega}{T^{r}_{m}/2} \frac{2 d}{c}+\omega{c} \frac{2 v}{c}\right)t_{d_{1}}
\end{aligned}
\end{equation}
for falling edge. The approximation holds for $\omega_{c}\gg\Delta\omega$, $T^{r}_{m}\gg\tau$ and $\sigma\rightarrow\infty$. The quantum beat signal has similar characteristics to those presented in the sawtooth modulation, except that the target's distance $d$ and velocity $v$ will be derived from the beat frequencies in each edge.
By analogy to the optimal detection strategy for $n$-photon NOON state \cite{lanzagorta2011synthesis}, the FMCW $n$-photon $(n\geq2)$ NOON state, Eq. (\ref{equ11}), with a detection strategy
\begin{equation}
\begin{aligned}
\hat{A}_{D}(t)&=-(|n_{t}, 0\rangle\langle 0, n_{t}|+| 0, n_{t}\rangle\langle n_{t}, 0|)\\
&=(|\psi_{D_{1}}(t)\rangle\langle \psi_{D_{1}}(t)|-| \psi_{D_{0}}(t)\rangle\langle \psi_{D_{0}}(t)|),
\end{aligned}
\end{equation}
where
\begin{equation}
|\psi_{D_{1}}(t)\rangle= \frac{1}{\sqrt{2}}(|n_{t}, 0\rangle-| 0,n_{t}\rangle),\quad |\psi_{D_{0}}(t)\rangle= \frac{1}{\sqrt{2}}(|n_{t}, 0\rangle+| 0,n_{t}\rangle),
\end{equation}
can provide $n$-fold improvement in detection resolution over the FMCW laser. Under this detection strategy, the normalized detection probability distributions in different times $t$ are
\begin{equation}
p_{n}(1|\vec{x},t)=T_{m}\langle \psi''_{n}(\vec{x})|\psi_{D_{1}}(t)\rangle \langle \psi_{D_{1}}(t) |  \psi''_{n} (\vec{x}) \rangle,
\end{equation}
and
\begin{equation}
p_{n}(0|\vec{x},t)=1-p_{n}(1|\vec{x},t)=T_{m}\langle \psi''_{n} (\vec{x})|\psi_{D_{0}}(t)\rangle \langle \psi_{D_{0}}(t) |  \psi''_{n} (\vec{x}) \rangle,
\end{equation}
which are the same as Eq. (\ref{equ12}) and Eq. (\ref{equ13}) with $n>2$ for triangular modulation.

\section{Precision limit of the quantum FMCW LiDAR}

In multi-parameter estimation theory, the Cramér-Rao bound (CRB) establishes a lower bound for the covariance matrix:
\begin{equation}
\operatorname{Cov}[\hat{x}] \geqslant \frac{1}{\nu} F^{-1} _{C}[\vec{x}]\geqslant \frac{1}{\nu} F^{-1} _{Q}[\vec{x}],
\end{equation}
here $\nu$ represents the number of times the procedure is repeated, $\vec{x}=(x_{1},x_{2},x_{3},\cdots)$ is a set of parameters and $\hat{x}=(\hat{x}_1,\hat{x}_2,\cdots)$ is the unbiased estimators of $\vec{x}$,
\begin{equation}
\operatorname{Cov}[\hat{x}]_{i j}=E[(\hat{x}_i-x_i)(\hat{x}_j-x_j)],
\end{equation}
 $F_{C}[\vec{x}]$ is the classical Fisher information (CFI) matrix, $F_{Q}[\vec{x}]$ is the quantum Fisher information (QFI) matrix. If $\operatorname{Cov}[\hat{x}]_{ij}=0$ for $i\neq j$, $x_{i}$ and $x_{j}$ are then uncorrelated.

For a quantum state $\rho(\vec{x})$ with unknown parameters $\vec{x}$, the elements of the QFI matrix is given by
\begin{equation}\label{equ19}
F_Q[\vec{x}]_{i j}=\operatorname{Tr}\left[\rho(\vec{x})\frac{L_{x_i} L_{x_j}+L_{x_j} L_{x_i}}{2}\right]
\end{equation}
where $L_{x_i}$ is the symmetric logarithmic derivative (SLD) associated with the parameter $x_i$, satisfying
\begin{equation}
\partial_{x_i} \rho(\vec{x})=\frac{1}{2}\left[\rho(\vec{x}) L_{x_i}+L_{x_i} \rho(\vec{x})\right].
\end{equation}
If $\rho(\vec{x})=\sum_l p_l\left|e_l\right\rangle\left\langle e_l\right|$ which is diagonal under the set of $k$-dimensional basis vectors $\{|e_{1}\rangle,|e_{2}\rangle,|e_{3}\rangle,\ldots,|e_{k}\rangle\}$, the SLD of $x_i$ is given by \cite{huangQuantumLimitedEstimationRange2021}
\begin{equation}\label{equ16}
L_{x_{i}}=2 \sum_{l, m: p_l+p_m \neq 0} \frac{\left\langle e_m\left|\partial_{x_{i}} \hat{\rho}(\vec{x})\right| e_l\right\rangle}{p_l+p_m}\left|e_m\right\rangle\left\langle e_l\right|.
\end{equation}
Note that, due to the uncertainty relation of non-commutative observables, it is not always possible to achieve the quantum CRB of two parameters simultaneously. A necessary and sufficient condition for two parameters $\{x_i,x_j\}$ to simultaneously achieve the simultaneous achieve the quantum CRB is
\begin{equation}
\operatorname{Tr}\left\{\rho(\vec{x}) \left[L_{x_i}, L_{x_j}\right]\right\}=0,
\end{equation}
which is the weak commutative condition.

In particular, for a pure state $\rho(\vec{x})=|\psi(\vec{x})\rangle\langle\psi(\vec{x})|$ with unknown parameter $\vec{x}$, 
\begin{equation}\label{equ17}
\partial_{x_i} \rho(\vec{x})=|\partial_{x_i}\psi(\vec{x})\rangle\langle\psi(\vec{x})|+|\psi(\vec{x})\rangle\langle\partial_{x_i}\psi(\vec{x})|, 
\end{equation}
and the element of QFI matrix will be reduced to 
\begin{equation}
F_{Q}[\vec{x}]_{i, j}=4 \operatorname{Re}(\left\langle\partial_{x_i} \psi(\vec{x}) \mid \partial_{x_j} \psi(\vec{x})\right\rangle-\left\langle\partial_{x_i} \psi(\vec{x}) \mid \psi(\vec{x})\right\rangle\langle\psi(\vec{x}) \mid \partial_{x_j} \psi(\vec{x})\rangle).
\end{equation}

On the other hand, by choosing a specific set of positive operator valued measurements (POVM) $\{ \Pi_{y} \}$, the probability of the measurement outcome $y$ is given by
\begin{equation}
p(y|\vec{x})=\langle \psi(\vec{x}) |\Pi_{y} |\psi(\vec{x})\rangle.
\end{equation}
Under this measurement, the CFI matrix can be obtained as
\begin{equation}
F_{C}[\vec{x}]_{i, j}=-\sum_{y}p(y | \vec{x})\frac{\partial^{2} \ln p(y | \vec{x})} {\partial x_{i} \partial x_{j} }.
\end{equation}

The CFI matrix of multiple independent experiments is just the sum of the CFI matrix for each experiment,  even if the probability distribution for each experiment is different,
\begin{equation}
F_{C}[\vec{x}]_{i, j}=-\sum_{n}\sum_{y}p_{n}(y | \vec{x})\frac{\partial^{2} \ln p_{n}(y | \vec{x})} {\partial x_{i} \partial x_{j} }
\end{equation}
here $p_{n}(y | \vec{x})$ is the probability distribution for the $n$-th experiment.

\subsection{Precision limit of quantum FMCW LiDAR under the sawtooth modulation}
 
The set of parameter $\vec{x}$ concerned in the sawtooth modulation are the target's relative distance $d$, and relative phase $\theta$, i.e., $\vec{x}=\{d,\theta\}$. As $\omega_{b}=(\Delta\omega/T^{s}_{m})(2d/c)$, here $T^{s}_{m}$ denotes the period of the sawtooth modulation, is a linear function of $d$, the precision limit can also be calibrated with $\vec{x}=\{\omega_{b},\theta\}$ instead of $\vec{x}=\{d,\theta\}\}$.

For the sawtooth FMCW $n$-photon NOON state $|\psi''_n(\vec{x})\rangle$ in Eq. (\ref{equ11}), the $ij$-th entry of the QFI matrix is
\begin{equation}
\begin{aligned}
F_{Q}\left[\omega_b,\theta\right]_{i,j}=&\frac{2}{T^{s}_{m}} \int^{t_{0}+T^{s}_{m}}_{t_{0}} d t\left[\partial_{x_{i}} E_{\omega_{0}}^{*n}\left(\omega_b,\theta,t\right)\right]\left[\partial_{x_{j}} E_{\omega_{0}}^{n}\left(\omega_b,\theta,t\right)\right]-\\
&\left[\frac{1}{T^{s}_{m}} \int^{t{0}+T^{s}_{m}}_{t_{0}} d t E_{\omega_{0}}^{n}\left(\omega_b,\theta,t\right) \partial_{x_{j}} E_{\omega_{0}}^{*n}\left(\omega_b,\theta,t\right)\right]\left[\frac{1}{T^{s}_{m}} \int^{t_{0}+T^{s}_{m}}_{t_{0}} d t E_{\omega_{0}}^{*n}\left(\omega_b,\theta,t\right) \partial_{x_{i}} E_{\omega_{0}}^{n}\left(\omega_b,\theta,t\right)\right]
\end{aligned}
\end{equation}
where
\begin{equation}
\begin{aligned}
E_{\omega_{0}}\left(\omega_b,\theta,t\right)\equiv E_{\omega_{0}}\left(t-\frac{2d}{c}\right)&=\exp \left\{i\left[\omega_{0} (t-t_{0}-\frac{2d}{c})+\frac{\Delta \omega }{2 T^{s}_{m}}(t-t_{0}-\frac{2d}{c})^{2}\right]\right\}\\
&\approx\exp \left\{i\left[\omega_{0} (t-t_{0})+\frac{\Delta \omega }{2 T^{s}_{m}}(t-t_{0})^{2}-\frac{\Delta \omega }{T^{s}_{m}}\frac{2d}{c}t-(\omega_{0} -\frac{\Delta \omega}{ T^{s}_{m}}t_{0})\frac{2d}{c}\right]\right\} \\
&=\exp \left\{i\left[\omega_{0} (t-t_{0})+\frac{\Delta \omega }{2 T^{s}_{m}}(t-t_{0})^{2}-\omega_{b}t-\theta\right]\right\},
\end{aligned}
\end{equation}
$\omega_{b}=\frac{\Delta \omega}{T_{m}^{s}} \frac{2 d}{c}$, $\theta=\left(\omega_{0}-\frac{\Delta \omega}{T_{m}^{s}} t_{0}\right) \frac{2 d}{c},$
and the approximation holds for $T^{s}_m>>\tau=2d/c$.
The QFI matrix is then 
\begin{equation}
\begin{aligned}
F_{Q}[\left\{\omega_b,\theta\right\}]=
\begin{pmatrix}
\frac{5 n^{2} T_m^{s2}}{12}+n^{2} t_0(t_0+T_m) & \quad \frac{n^{2} T^{s}_m}{2}+n^{2} t_0 \\ \\
\frac{n^{2} T^{s}_m}{2}+n^{2} t_0 & \quad n^{2}
\end{pmatrix},
\end{aligned}
\end{equation}
which depends on the initial modulation time $t_{0}$. If the photon source emits $N$ FMCW $n$-photon NOON states per unit time, then for a single modulation period $T_m$ with $\nu$ times repetition of the emission, the quantum CRB of $\vec{x}=\{\omega_{b},\theta\}$ is given by:
\begin{equation}
\begin{aligned}
\operatorname{Cov}(\left\{\omega_b,\theta\right\})\geq \frac{1}{N \nu T^{s}_{m}}F^{-1}_{Q}[\left\{\omega_b,\theta\right\}]= \frac{1}{N \nu  T^{s}_m}
\begin{pmatrix}
\frac{6}{n^{2} T_m^{s2}} &\quad -\frac{3}{n^{2} T^{s}_m}-\frac{6  t_0}{n^{2} T_m^{s2}} \\ \\
-\frac{3}{n^{2} T^{s}_m}-\frac{6 t_0}{n^{2} T_m^{s2}} & \quad  \frac{5}{2 n^{2}}+\frac{6 t_0(t_0+T^{s}_m)}{n^{2} T_m^{s2}}
\end{pmatrix}.
\end{aligned}
\end{equation}
It can be seen that there is a correlation between $\omega_{b}$ and $\theta$, which affects the precision limit of $\theta$. To remove the correlation, we can choose $t_{0}=-T^{s}_{m}/2$ so that
\begin{equation}
\begin{aligned}
\operatorname{Cov}(\left\{\omega_b,\theta\right\})\geq  \frac{1}{N \nu T^{s}_m}
\begin{pmatrix}
\frac{6}{n^{2} T_m^{s2}} &\quad 0 \\ \\
0 & \quad  \frac{1}{n^{2}}
\end{pmatrix}.
\end{aligned}
\end{equation}
In this case $t=0$ is the center time of single modulation period. 
On the other hand, under the detection strategy presented in Sec. \ref{IID}, the quantum beat signal with the $n$-photon FMCW NOON state under the sawtooth modulation is given by
\begin{equation}\
p_{n}\left(1 \mid t, \omega_{b}, \theta\right) \approx \frac{1}{2}\left[1+\cos \left(n\omega_{b} t+n \theta\right)\right]
\end{equation}
and $p_{n}\left(0 \mid t, \omega_{b}, \theta\right)=1-p_{n}\left(1 \mid t, \omega_{b}, \theta\right)$, here
$\theta=(\omega_{0}-\frac{\Delta \omega}{T_{m}^s}t_{d}) \frac{2d}{c}.$
The CFI matrix can then be obtained as 
\begin{equation}
F_{C}[\omega_b,\theta, t]_{i,j}=-p(0 \mid t, \omega_{b}, \theta)\frac{\partial^{2} \ln p(0 \mid t, \omega_{b}, \theta)} {\partial x_{i} \partial x_{j} }-p(1 \mid t, \omega_{b}, \theta)\frac{\partial^{2} \ln p(1 \mid t, \omega_{b}, \theta)} {\partial x_{i} \partial x_{j} }.
\end{equation}
Due to the additivity of CFI matrix, with $\nu$ repetitions of the detection over a period, the CFI matrix is
\begin{equation}
\begin{aligned}
F_{C}[\omega_b,\theta]=N \nu\int^{t_d+T^{s}_{m}}_{t_d}dt F_{C}[\omega_b,\theta,t]= N \nu T^{s}_{m}
\begin{pmatrix}
\frac{n^{2} T_m^{s2}}{3}+n^{2} t_d(t_d+T^{s}_m) & \quad\frac{n^{2} T^{s}_m}{2}+n^{2} t_d \\ \\
\frac{n^{2} T^{s}_m}{2}+n^{2} t_d & \quad n^{2}
\end{pmatrix},
\end{aligned}
\end{equation}
where $t_d$ is the initial time. In this case, the classical CRB is given by
\begin{equation}
\begin{aligned}
\operatorname{Cov}(\left\{\omega_b,\theta\right\})\geq F^{-1}_{C}[\left\{\omega_b,\theta\right\}]= \frac{1}{N \nu  T^{s}_m}
\begin{pmatrix}
\frac{12}{n^{2} T_m^{s2}} &\quad -\frac{6}{n^{2} T^{s}_m}-\frac{12 t_d}{n^{2} T_m^{s2}} \\ \\
-\frac{6}{n^{2} T^{s}_m}-\frac{12 t_d}{n^{2} T_m^{s2}} &\quad \frac{4}{n^{2}}+\frac{12 t_d(t_d+T^{s}_m)}{n^{2} T_m^{s2}}
\end{pmatrix}.
\end{aligned}
\end{equation}
The correlation between $\omega_{b}$ and $\theta$ can be removed by choosing $t_d=-T^{s}_{m}/2$, which leads to
\begin{equation}
\begin{aligned}
\operatorname{Cov}(\left\{\omega_b,\theta\right\})\geq  \frac{1}{N \nu  T^{s}_m}
\begin{pmatrix}
\frac{12}{n^{2} T_m^{s2}} &\quad 0 \\ \\
0 & \quad  \frac{1}{n^{2}}
\end{pmatrix},
\end{aligned}
\end{equation}

In summary, with $t_{0}=t_d=-T^{s}_{m}/2$ the precision limit is given by
\begin{equation}
\begin{aligned}
\operatorname{Cov}(\left\{d,\theta\right\})\geq  \frac{1}{N \nu T^{s}_m}
\begin{pmatrix}
\frac{3 }{2n^{2}  }\frac{c^{2}}{\Delta\omega^{2}} &\quad 0 \\ \\
0 & \quad  \frac{1}{n^{2}}
\end{pmatrix}.
\end{aligned}
\end{equation}
The comparison between entangled biphoton states ($n=2$, $N/2$) and separable photon states ($n=1$, $N$) shows that using entangled states achieve better precision.

\subsection{Precision limit under the triangular modulation\label{IIIB}}

The set of parameters involved in the triangular modulation are target’s distance $d$, target’s velocity $v$, relative phase $\theta_{0}$ in the rising edge, and relative phase in the falling edge $\theta_{1}$, i.e., $\vec{x}=\{d,v,\theta_{0},\theta_{1}\}$. Since the sum and difference of $\omega_{b_{1}}$ and $|\omega_{b_{2}}|$ are linear functions of $d$ and $v$,
\begin{equation}
\omega_{b}\equiv\frac{\omega_{b_{1}}+|\omega_{b_{2}}|}{2} =\frac{\Delta \omega }{T_{m}^{r}/2 }\frac{2d}{c}, \quad \omega_{d}=\frac{\omega_{b_{1}}-|\omega_{b_{2}}|}{2} = \omega_{c}\frac{2v}{c},
\end{equation}
where $T_m^r$ is the period for the triangular modulation. We can thus calculate the precision limit of $\vec{x}=\{\omega_{b},\omega_{d},\theta_{0},\theta_{1}\}$ instead.

For $n$-photon FMCW NOON state under the triangular modulation, as $|\psi''_{n}(\vec{x})\rangle$ in Eq. (\ref{equ11}), the QFI matrix can be obtained as
\begin{equation}
\begin{aligned}
F_{Q}\left[\left\{\omega_{b},\omega_{d},\theta_{0},\theta_{1}\right\}\right]_{i,j}=&\frac{2}{T^{r}_{m}} \int^{t’_{0}+T^{r}_{m}/2}_{t’_{0}-T^{r}_{m}/2} d t\left[\partial_{x_{i}} E_{\omega_{0}}^{*n}\left(\omega_{b},\omega_{d},\theta_{0},\theta_{1},t\right)\right]\left[\partial_{x_{j}} E_{\omega_{0}}^{n}\left(\omega_{b},\omega_{d},\theta_{0},\theta_{1},t\right)\right]-\\
&\left[\frac{1}{T^{r}_{m}} \int^{t’_{0}+T^{r}_{m}/2}_{t’_{0}-T^{r}_{m}/2} d t E_{\omega_{0}}^{n}\left(\omega_{b},\omega_{d},\theta_{0},\theta_{1},t\right) \partial_{x_{j}} E_{\omega_{0}}^{*n}\left(\omega_{b},\omega_{d},\theta_{0},\theta_{1},t\right)\right] \times\\
&\left[\frac{1}{T^{r}_{m}} \int^{t’_{0}+T^{r}_{m}/2}_{t’_{0}-T^{r}_{m}/2} d t E_{\omega_{0}}^{*n}\left(\omega_{b},\omega_{d},\theta_{0},\theta_{1},t\right) \partial_{x_{i}} E_{\omega_{0}}^{n}\left(\omega_{b},\omega_{d},\theta_{0},\theta_{1},t\right)\right]
\end{aligned}
\end{equation}
where at the rising edge $t\in[t’_{0}-T^{r}_{m}/2,t’_{0}]$, 
\begin{equation}\label{equ14}
\begin{aligned}
&E_{\omega_{0}}\left(\omega_{b},\omega_{d},\theta_{0},t\right) \equiv E_{\omega_{0}-\omega_d}\left(t-\frac{2d}{c}\right)\\
&=\exp \left\{-i\left[\left(\omega_0-\omega_d+\Delta \omega\right)\left(t-\frac{2 d}{c}-t_0^{\prime}\right)+\frac{\Delta \omega}{T_m^r}\left(t-\frac{2 d}{c}-t_0^{\prime}\right)^2\right]\right\}\\
&\approx\exp \left\{i[-(\omega_{0}+\Delta \omega)(t-t'_{0})-\frac{\Delta \omega }{ T^{r}_{m}}(t-t'_{0})^{2}+(\omega_{b}+\omega_{d})(t-t'_{0})+(\omega_{0}-\omega_{d}+\Delta \omega) \frac{2d}{c}]\right\}\\
&=\exp \left\{i[-(\omega_{0}+\Delta \omega)(t-t'_{0})-\frac{\Delta \omega }{ T^{r}_{m}}(t-t'_{0})^{2}+(\omega_{b}+\omega_{d})(t-t’_{0}+T_{1})+\theta_{0}]\right\}
\end{aligned}
\end{equation}
with
$\theta_{0}=(\omega_{0}-\omega_{d}+\Delta \omega) \frac{2d}{c}-(\omega_{b}+\omega_{d})T_{1}$ which is the phase of beating signal at time $t=t’_{0}-T_{1}$,
and at the falling edge, $t\in[t’_{0}, t’_{0}+T^{r}_{m}/2]$,
\begin{equation}\label{equ15}
\begin{aligned}
&E_{\omega_{0}}\left(\omega_{b},\omega_{d},\theta_{1},t\right) \equiv E_{\omega_{0}-\omega_d}\left(t-\frac{2d}{c}\right)\\
&\approx\exp \left\{i[(\omega_{0}+\Delta \omega)(t-t_{0})-\frac{\Delta \omega }{ T^{r}_{m}}(t-t_{0})^{2}-(-\omega_{b}+\omega_{d})(t-t_{0})-(\omega_{0}-\omega_{d}+\Delta \omega)  \frac{2d}{c}]\right\}\\
&=\exp \left\{i[(\omega_{0}+\Delta \omega)(t-t_{0})-\frac{\Delta \omega }{ T^{r}_{m}}(t-t_{0})^{2}-(-\omega_{b}+\omega_{d})(t-t_{0}-T_{2})-\theta_{1}]\right\}
\end{aligned}
\end{equation}
with
$\theta_{1}=(\omega_{0}-\omega_{d}+\Delta \omega)  \frac{2d}{c}+(-\omega_{b}+\omega_{d})T_{2}$ which is the phase of beating signal at time $t=t’_{0}+T_{2}$,
here the approximation holds for $T^{r}_{m}>>\tau=2d/c$. 

Similar to the case of the sawtooth modulation, the presence of $\theta_{0}$ and $\theta_{1}$ may affect the precision through the correlations between the measurement of the frequency and phase. Such correlation under the triangular modulation, however, can be removed by choosing $T_{1}=T_{2}=T^{r}_{m}/4$, under which the QFI matrix becomes
\begin{equation}
\begin{aligned}
F_{Q}[\left\{\omega_{b},\omega_{d},\theta_{0},\theta_{1}\right\}]= \begin{pmatrix}
\frac{n^{2} T_m^{r2}}{24} & 0 & 0 & 0 \\ \\
0 & \frac{n^{2} T_m^{r2}}{24} & 0 & 0 \\ \\
0 & 0 & \frac{3 n^{2}}{4} & \frac{n^{2}}{4} \\ \\
0 & 0 & \frac{n^{2}}{4} & \frac{3 n^{2}}{4}
\end{pmatrix}.
\end{aligned}
\end{equation}
For a single modulation period $T^{r}_{m}$, if the photon source emits an FMCW $n$-photon NOON state per unit time, and the emission is repeated $\nu$ times during the period $T^{r}_m$, then the quantum Cramér-Rao bound (CRB) is given by:
\begin{equation}
\begin{aligned}
\operatorname{Cov}(\left\{\omega_{b},\omega_{d},\theta_{0},\theta_{1}\right\})\geq \frac{1}{\nu T^{r}_{m}}F^{-1}_{Q}[\left\{\omega_{b},\omega_{d},\theta_{0},\theta_{1}\right\}]
= \frac{1}{\nu  T^{r}_m} 
\begin{pmatrix}
\frac{24}{n^{2} T_m^{r2}} & 0 & 0 & 0 \\ \\
0 & \frac{24}{n^{2} T_m^{r2}} & 0 & 0 \\ \\
0 & 0 & \frac{3}{2 n^{2}} & -\frac{1}{2 n^{2}} \\ \\
0 & 0 & -\frac{1}{2 n^{2}} & \frac{3}{2 n^{2}}.
\end{pmatrix}
\end{aligned}
\end{equation}
 The quantum CRB of $\vec{x}=\{d,v,\theta_{0},\theta_{1}\}$ is
\begin{equation}\label{equ20}
\begin{aligned}
\operatorname{Cov}(\left\{d,v,\theta_{0},\theta_{1}\right\})\geq  \frac{1}{\nu T^{r}_{m}}
\begin{pmatrix}
\frac{3}{2n^{2}} \frac{c^{2}}{\Delta \omega^{2}}  & 0 & 0 & 0 \\ \\
0 & \frac{6}{n^{2}} \frac{c^{2}}{\omega_{c}^{2} T_m^{r2}}  & 0 & 0 \\ \\
0 & 0 & \frac{3}{2 n^{2}} & -\frac{1}{2 n^{2}} \\ \\
0 & 0 & -\frac{1}{2 n^{2}} & \frac{3}{2 n^{2}}
\end{pmatrix}.
\end{aligned}
\end{equation}
From which we can obtain the quantum CRB for the time delay, $\delta\tau$, and the Doppler shift, $\delta\omega_d$, as(assume $\nu T^{r}_{m}=1$)
\begin{equation}
\delta\tau\geq\frac{\sqrt{6}}{ n\Delta\omega}, \quad \delta\omega_d\geq\frac{2\sqrt{6}}{ nT^{r}_{m}}.
\end{equation} 
which is derived from Eq.(\ref{equ23}-\ref{equ24}) and (\ref{equ20}).

On the other hand, according to the detection strategy presented in Section \ref{IID}, the quantum beat signals obtained from $n$-photon FMCW NOON state under the triangular modulation is 
\begin{equation}\label{eq:FMCWbeat1}
\begin{aligned}
&p_{n}(1 \mid \omega_{b},\omega_{d}, \theta_{0},\theta_{1}, t) \approx  \frac{1}{2}\left\{1+\cos \left[n\left(\omega_{b}+\omega_{d}\right) t+n \theta_{0} \right]\right\},
\end{aligned}
\end{equation}
with $\theta_{1}=0$, 
$\theta_{0}= (\omega_{0}-\omega_d+\Delta \omega) \frac{2 d}{c}-\left(\omega_{b}+\omega_d\right)t_{d_{0}}$ 
for $t\in[t_{d_{0}}-T^{r}_{m}/2, t_{d_{0}}]$ and an initial detection time $t_{d_{0}}-T^{r}_{m}/2$, and for $t\in[t_{d_{1}},t_{d_{1}}+T^{r}_{m}/2]$ with an initial detection time $t_{d_{1}}$
\begin{equation}\label{eq:FMCWbeat2}
\begin{aligned}
&p_{n}(1 \mid \omega_{b},\omega_{d}, \theta_{0},\theta_{1}, t) \approx \frac{1}{2}\left\{1+\cos \left[n\left(-\omega_{b}+\omega_{d}\right) t+n \theta_{1}\right]\right\},
\end{aligned}
\end{equation}
with $\theta_{0}=0$, 
$\theta_{1}= (\omega_{0}-\omega_d+\Delta \omega) \frac{2 d}{c}-\left(-\omega_{b} +\omega_{d}\right)t_{d_{1}}$. 
 And
\begin{equation}
p_{n}(0 \mid \omega_{b},\omega_{d}, \theta_{0},\theta_{1}, t)=1-p_{n}(1 \mid \omega_{b},\omega_{d}, \theta_{0},\theta_{1}, t).
\end{equation}
 By choosing $t_{d_{0}}=-t_{d_{1}}=T^{r}_{m}/4$, the correlation between the frequency and phase measurement in the classical CRB will be decouple as we will show below. 

Due to the additivity of the CFI matrix $F_{C}\left[\left\{\omega_b,\omega_d,\theta_{0},\theta_{1}\right\}\right|t]$ for each measurement, the CFI matrix for a single modulation period $T_m$ with $\nu$ repetitions of the detection can be obtained by summing the CFI matrices for each measurement. Thus, the CFI matrix for the total detection time is given by
\begin{equation}
\begin{aligned}
F_{C}\left[\omega_b,\omega_d,\theta_{0},\theta_{1}\right]&=\nu\int^{t_{d_{0}}}_{t_{d_{0}}-T^{r}_{m}/2}dt F_{C}\left[\left\{\omega_b,\omega_d,\theta_{0},\theta_{1}\right\}\right|t]+\nu\int^{t_{d_{1}}+T^{r}_{m}/2}_{t_{d_{1}}} dt F_{C}\left[\left\{\omega_b,\omega_d,\theta_{0},\theta_{1}\right\}\right|t]\\
&=\nu T^{r}_{m}
\begin{pmatrix}
\frac{n^{2} T_m^{r2}}{48} & 0 & 0 & 0 \\ \\
0 & \frac{n^{2} T_m^{r2}}{48} & 0 & 0 \\ \\
0 & 0 & \frac{n^{2}}{2} & 0 \\ \\
0 & 0 & 0 & \frac{n^{2}}{2}
\end{pmatrix}.
\end{aligned}
\end{equation}
The classical CRB is then given by
\begin{equation}
\begin{aligned}
\operatorname{Cov}(\left\{\omega_{b},\omega_{d},\theta_{0},\theta_{1}\right\})\geq F^{-1}_{C}[\left\{\omega_{b},\omega_{d},\theta_{0},\theta_{1}\right\}]
= \frac{1}{\nu  T^{r}_m} 
\begin{pmatrix}
\frac{48}{n^{2} T_m^{r2}} & 0 & 0 & 0 \\ \\
0 & \frac{48}{n^{2} T_m^{r2}} & 0 & 0 \\ \\
0 & 0 & \frac{2}{ n^{2}} & 0 \\ \\
0 & 0 & 0 & \frac{2}{ n^{2}}
\end{pmatrix}.
\end{aligned}
\end{equation}

From which we can also obtain the classical CRB of $\vec{x}=\{d,v,\theta_{0},\theta_{1}\}$ as
\begin{equation}\label{equ25}
\begin{aligned}
\operatorname{Cov}(\left\{d,v,\theta_{0},\theta_{1}\right\})\geq  \frac{1}{\nu T^{r}_{m}}
\begin{pmatrix}
\frac{3}{n^{2}} \frac{c^{2}}{\Delta \omega^{2}}  & 0 & 0 & 0 \\ \\
0 & \frac{12}{n^{2}} \frac{c^{2}}{\omega_{c}^{2} T_m^{r2}}  & 0 & 0 \\ \\
0 & 0 & \frac{2}{n^{2}}  & 0 \\ \\
0 & 0 & 0 &  \frac{2}{n^{2}},
\end{pmatrix}
\end{aligned}
\end{equation}
where we have chosen $t_{d_{0}}=-t_{d_{1}}=T^{r}_{m}/4$. 
The classical CRB for the time delay $\delta\tau$ and the Doppler frequency shift, $\delta\omega_d$, are given by( with $\nu T^{r}_{m}=1$)
\begin{equation}
\delta\tau\geq\frac{2\sqrt{3}}{ n\Delta\omega}, \quad \delta\omega_d\geq\frac{4\sqrt{3}}{ nT^{r}_{m}},
\end{equation}
which can be easily obtained from Eq.(\ref{equ23}-\ref{equ24}) and (\ref{equ25}).

\subsection{The saturation of the quantum CRB for the estimation of $\{d,v\}$}

In this subsection, we restrict our discussion to simultaneous ranging and velocity measurement $\vec{x}=\{d,v\}$ and show the weak commutative condition for the state $\rho(\vec{x})=|\psi''_{n}(\vec{x})\rangle\langle\psi''_{n}(\vec{x})|$
\begin{equation}
\operatorname{Tr}\left\{\hat{\rho}(\vec{x}) \left[L_{d}, L_{v}\right]\right\}=0,
\end{equation}
is satisfied, where 
\begin{equation}
\begin{aligned}
\left|\psi''_{n}(\vec{x})\right\rangle=\frac{1}{\sqrt{ 2T^{r}_{m}}} \int^{t_{0}+T^{r}_{m}}_{t_{0}} d t&\left\{E_{\omega_{c}}^{* n}(t)\left|n_{t}, 0\right\rangle-E_{\omega_{c}-\omega_{d}}^{* n}(t-\tau) \left|0,n_{t}\right\rangle\right\},
\end{aligned}
\end{equation}
is the $n$-photon triangular FMCW NOON state when $j_{B}\rightarrow\infty$ and $T_{1}=T_{2}=T^{r}_{m}/4$.

To obtain the SLD for $\vec{x}=\{d,v\}$ through Eq. (\ref{equ16}), we let 
$\{|e_{1}\rangle,|e_{2}\rangle,|e_{3}\rangle\}$ where $|e_{1}\rangle\equiv|\psi''_{n}(\vec{x})\rangle$,
\begin{equation}
\begin{aligned}
|e_{2}\rangle\equiv a\left|\partial_{d}\psi''_{n}(\vec{x})\right\rangle=-\frac{a}{\sqrt{ 2T^{r}_{m}}} \int^{t_{0}+T^{r}_{m}}_{t_{0}} d t [\partial_{d}E_{\omega_{c}-\omega_{d}}^{* n}(t-\tau)] \left|0,n_{t}\right\rangle,
\end{aligned}
\end{equation}
\begin{equation}
\begin{aligned}
|e_{3}\rangle\equiv b\left|\partial_{v}\psi''_{n}(\vec{x})\right\rangle=-\frac{b}{\sqrt{ 2T^{r}_{m}}} \int^{t_{0}+T^{r}_{m}}_{t_{0}} d t [\partial_{v}E_{\omega_{c}-\omega_{d}}^{* n}(t-\tau)] \left|0,n_{t}\right\rangle,
\end{aligned}
\end{equation}
where $a=\frac{\sqrt{6}}{n}\frac{c}{\Delta \omega}, b=\frac{2 \sqrt{6} }{n  }\frac{c}{\omega_{c}T^{r}_{m}}$
are the normalization factors. It is easy to verify that these three vectors are mutually orthogonal. 
If we use these vectors as a basis for a 3-dimensional space, we can write the state $\rho(\vec{x})=|\psi''_{n}(\vec{x})\rangle\langle\psi''_{n}(\vec{x})|$ as
\begin{equation}\label{equ18}
\begin{aligned}
\rho(\vec{x})= 
\begin{pmatrix}
1 \quad\quad& 0 \quad\quad& 0  \\ \\
0 \quad\quad& 0 \quad\quad& 0  \\ \\
0 \quad\quad& 0 \quad\quad& 0  
\end{pmatrix}.
\end{aligned}
\end{equation}

The SLDs of $\vec{x}=\{d,v\}$, $L_{d}$ and $L_{v}$, can be obtained as
\begin{equation}\label{equ21}
\begin{aligned}
\hat{L}_{d}= 
\begin{pmatrix}
0 \quad\quad& 2/a \quad\quad& 0  \\ \\
2/a \quad\quad& 0 \quad\quad& 0  \\ \\
0 \quad\quad& 0 \quad\quad& 0   
\end{pmatrix}= 
\begin{pmatrix}
0 \quad\quad& \frac{2  n  }{\sqrt{6}  }\frac{\Delta \omega}{c} \quad\quad& 0  \\ \\
\frac{2  n  }{\sqrt{6}  }\frac{\Delta \omega}{c} \quad\quad& 0 \quad\quad& 0  \\ \\
0 \quad\quad& 0 \quad\quad& 0  
\end{pmatrix},
\end{aligned}
\end{equation}
\begin{equation}\label{equ22}
\begin{aligned}
\hat{L}_{v}= 
\begin{pmatrix}
0 \quad\quad& 0 \quad\quad& 2/b  \\ \\
0 \quad\quad& 0 \quad\quad& 0  \\ \\
2/b \quad\quad& 0 \quad\quad& 0   \\ \\
\end{pmatrix}= 
\begin{pmatrix}
0 \quad\quad& 0 \quad\quad& \frac{n  }{\sqrt{6} }\frac{\omega_{c}T^{r}_{m}}{c}  \\ \\
0 \quad\quad& 0 \quad\quad& 0  \\ \\
\frac{n  }{\sqrt{6} }\frac{\omega_{c}T^{r}_{m}}{c} \quad\quad& 0 \quad\quad& 0  
\end{pmatrix}.
\end{aligned}
\end{equation}
And the QFI matrix is
\begin{equation}
\begin{aligned}
F_{Q}[\left\{d,v\right\}]= 
\begin{pmatrix}
\frac{2n^{2}}{3} \frac{\Delta \omega^{2}}{c^{2}}  & 0  \\ \\
0 & \frac{n^{2}}{6} \frac{\omega_{c}^{2} T_m^{r2}}{c^{2}}   
\end{pmatrix},
\end{aligned}
\end{equation}
which gives the quantum CRB as
\begin{equation}
\begin{aligned}
\operatorname{Cov}(\left\{d,v\right\})\geq \frac{1}{\nu T^{r}_{m}}F^{-1}_{Q}[\left\{d,v\right\}]
= \frac{1}{\nu  T^{r}_m} 
\begin{pmatrix}
\frac{3}{2n^{2}} \frac{c^{2}}{\Delta \omega^{2}}  & 0  \\ \\
0 & \frac{6}{n^{2}} \frac{c^{2}}{\omega_{c}^{2} T_m^{r2}}   
\end{pmatrix},
\end{aligned}
\end{equation}
where we assume the measurement is repeated $\nu$ times in a unit time and the precision is calibrated for a single modulation period $T_m$. 

The weak commutative condition for the estimation of $\{d,v\}$ 
\begin{equation}
\operatorname{Tr}\left\{\rho(\vec{x}) \left[L_{d}, L_{v}\right]\right\}=0
\end{equation}
is satisified in this case, the quantum CRB is thus achievable. In particular, it can be achieved with the projective measurement, $\{|e_1\rangle\langle e_1|,|e_2\rangle\langle e_2|,|e_3\rangle\langle e_3|\}$. The optimal measurement, however, can be challenging in practise. The classical CRB under the coincidence counting is then typically used to calibrate the precision in practise.

\subsection{Resolution of the quantum FMCW LiDAR}

In quantum FMCW LiDAR, the discrete Fourier transform (DFT) of the observed quantum beat signal determines the resolution of $d,v$. The maximum-likelihood (ML) estimators \cite{rifeSingleToneParameter1974, erkmenMaximumlikelihoodEstimationFrequencymodulated2013} is equivalent to the DFT, and the classical CR bound can be attained using these ML estimators. As a result, the frequency resolution in the DFT controls the resolutions of $d,v$.

The DFT of the beat signal in Equation (\ref{equ26}) for sawtooth modulation can be obtained in one $T^s_m$ modulation cycle. This makes it possible to identify the angular frequency of the beat signal separated by $2\pi/T^s_m$:
\begin{equation}
n\frac{\Delta \omega}{T^{s}_{m}} \frac{2 \Delta d}{c}=\frac{2 \pi}{ T^{s}_{m}}
\end{equation}
Here, $\Delta d$ stands for the smallest distance that the quantum FMCW LiDAR can distinguish, or the resolution of the distance.

Because of this, the quantum FMCW LiDAR's resolution of $d$ is 
\begin{equation}
\Delta d=\frac{2 \pi}{ n} \frac{c}{2 \Delta \omega},
\end{equation}
which represents an improvement of $n$ in comparison to the traditional FMCW LIDAR with sawtooth modulation.

The DFT of the beat signals in Eq. (\ref{equ12}) and (\ref{equ13}) for triangular modulation can be obtained for each half period. This makes it possible to discriminate between the angular frequency of the beat signal separated by $\frac{2\pi}{T^r_m/2}$:
\begin{equation}
n\frac{\Delta \omega}{T^{r}_{m}/2} \frac{2 \Delta d}{c}=\frac{2 \pi}{ T^{r}_{m}/2}, \quad n\omega_{c} \frac{2 \Delta v}{c}=\frac{2 \pi}{ T^{r}_{m}/2}
\end{equation}
Here, $Delta v$, or the resolution of $v$, is the lowest velocity that the quantum FMCW LiDAR can distinguish. The quantum FMCW LiDAR's resolutions of $d,v$ are
\begin{equation}
\Delta d=\frac{2 \pi}{ n} \frac{c}{2 \Delta \omega}, \quad \Delta v=\frac{2 \pi}{n} \frac{c}{\omega_c  T^{r}_{m} }.
\end{equation}
When compared to traditional FMCW LIDAR with triangular modulation, this exhibits a factor of $n$ improvement in resolution.

\section{Precision limit and resolution of classical FMCW LIDAR}

For comparison, we also present the precision limit and resolution of classical FMCW Lidar with the coherent FMCW state under the discrete time and finite bandwidth approximation as
\begin{equation}
\begin{aligned}
\mathop{\otimes}_{j}\left|\alpha s_{\omega_{c}}^{*}\left(\omega_{j}\right)\right\rangle &=\prod^{\infty}_{j=-\infty} \exp \left[\alpha s_{\omega_{c}}^{*}\left(\omega_{j}\right) a^{+}\left(\omega_{j}\right)-\alpha^{*} s_{\omega_{c}}\left(\omega_{j}\right) a\left(\omega_{j}\right)\right]|0\rangle \\
&\approx\exp \left\{\sum^{j_{c}+\frac{j_{B}}{2}}_{j=j_{c}-\frac{j_{B}}{2}}\left[\alpha s_{\omega_{c}}^{*}\left(\omega_{j}\right) a^{+}\left(\omega_{j}\right)-\alpha^{*} s_{\omega_{c}}\left(\omega_{j}\right) a\left(\omega_{j}\right)\right]\right\}|0\rangle \\
& =\exp \left\{\frac{1}{\sqrt{j_{B}+1}} \sum^{p_{0}+j_{B}}_{p=p_{0}}\left[\alpha E_{\omega_{c}}^{*}\left(t_{p}\right) a^{+}\left(t_{p}\right)-\alpha^{*} E_{\omega_{c}}\left(t_{p}\right) a\left(t_{p}\right)\right]\right\}|0\rangle \\
&=\mathop{\otimes}^{p_{0}+j_{B}}_{p=p_{0}}\left|\frac{\alpha E^{*}\left(t_{p}\right)}{\sqrt{j_{B}+1}}\right\rangle,
\end{aligned}
\end{equation}
here 
\begin{equation}
a\left(t_{p}\right)\left|\frac{\alpha E_{\omega_{c}}^{*}\left(t_{p}\right)}{\sqrt{j_{B}+1}}\right\rangle=\frac{\alpha E_{\omega_{c}}^{*}\left(t_{p}\right)}{\sqrt{j_{B}+1}}\left|\frac{\alpha E_{\omega_{c}}^{*}\left(t_{p}\right)}{\sqrt{j_{B}+1}}\right\rangle
\end{equation}
is a coherent state in the discrete time domain similar to that of the single-mode coherent state $|\alpha\rangle$.

The state, $\left|\frac{\alpha E_{\omega_{c}}^{*}\left(t_{p}\right)}{\sqrt{j_{B}+1}}\right\rangle$, is characterized by a Poisson distribution of photon number fluctuations at time $t_p$, with an average value of $|\alpha|^2/(j_B+1)$, similar to that of a single-mode coherent state. As a result, the fluctuation of the state's light intensity at time $t_p$ satisfies a Poisson distribution, with an average value of $|\alpha|^2$.

After injecting the coherent FMCW  state $\mathop{\otimes}^{p_{0}+j_{B}}_{p=p_{0}}\left|\frac{\alpha E_{\omega_{c}}^{*}\left(t_{p}\right)}{\sqrt{j_{B}+1}}\right\rangle$ into port 1 of the MZ interferometer shown in Fig. \ref{fig:2}, we get
\begin{equation}
|\psi_{c}(\vec{x})\rangle=\mathop{\otimes}^{p_{0}+j_{B}}_{p=p_{0}}\left|\frac{\alpha }{2\sqrt{(j_{B}+1)}}[E_{\omega_{c}}^{*}\left(t_{p}\right)+E_{\omega_{c}-\omega_{d}}^{*}\left(t_{p}-\tau\right)]\right\rangle_{R’}\mathop{\otimes}^{p_{0}+j_{B}}_{p=p_{0}}\left|\frac{\alpha }{2\sqrt{(j_{B}+1)}}[E_{\omega_{c}}^{*}\left(t_{p}\right)-E_{\omega_{c}-\omega_{d}}^{*}\left(t_{p}-\tau\right)]\right\rangle_{T’}
\end{equation}
where $\vec{x}=\{d,v,\theta_{0},\theta_{1}\}$ (or $\{\omega_b,\omega_d,\theta_{0},\theta_{1}\}$ equivalently) at the port $R’$ and $T’$. 

In this section, we only consider the triangular frequency modulation for classical FMCW LiDAR. As shown in Figure \ref{fig:4} the average light intensity detected by PD1 can be calculated with the first-order quantum correlation function. For $t\in[t_{d_{0}}-T^{r}_{m}/2, t_{d_{0}}]$ with an initial detection time $t_{d_{0}}-T^{r}_{m}/2$, the average intensity is
\begin{equation}
\begin{aligned}
I_{c}(1| d,v, \theta_{0}, \theta_{1},t)&\equiv\langle\psi_{c}(d,v, \theta_{0}, \theta_{1})|\hat{E}_{R^{\prime}}^{(-)}\left(t\right) \hat{E}_{R^{\prime}}^{(+)}\left(t\right)|\psi_{c}(d,v, \theta_{0}, \theta_{1})\rangle\\
&=\left|\frac{\alpha }{2}[E_{\omega_{c}}^{*}\left(t\right)+E_{\omega_{c}-\omega_{d}}^{*}\left(t-\tau\right)]\right|^{2}\\
&\approx \frac{1}{2} |\alpha|^{2}\left\{1+\cos \left[\left(\omega_{b}+\omega_{d}\right) t+ \theta_{0} \right]\right\},
\end{aligned}
\end{equation}
here
\begin{equation}
\begin{aligned}
\omega_{b}=\frac{\Delta \omega }{T_{m}^{r}/2 }\frac{2d}{c}, \quad \omega_{d} = \omega_{c}\frac{2v}{c}, \quad\theta_{0}= (\omega_{0}-\omega_d+\Delta \omega) \frac{2 d}{c}-\left(\omega_{b}+\omega_d\right)t_{d_{0}},  \quad\theta_{1}=0,
\end{aligned}
\end{equation}
and for $t\in[t_{d_{1}},t_{d_{1}}+T^{r}_{m}/2]$ with an initial detection time $t_{d_{1}}$,
\begin{equation}
\begin{aligned}
&I_{c}(1 \mid d,v, \theta_{0}, \theta_{1},t) \approx \frac{1}{2}|\alpha|^{2}\left\{1+\cos \left[\left(-\omega_{b}+\omega_{d}\right) t+ \theta_{1}\right]\right\},
\end{aligned}
\end{equation}
here
\begin{equation}
\begin{aligned}
\omega_{b}=\frac{\Delta \omega }{T_{m}^{r}/2 }\frac{2d}{c}, \quad \omega_{d} = \omega_{c}\frac{2v}{c}, \quad\theta_{0}=0,\quad\theta_{1}= (\omega_{0}-\omega_d+\Delta \omega) \frac{2 d}{c}-\left(-\omega_{b} +\omega_{d}\right)t_{d_{1}} .
\end{aligned}
\end{equation}
And the average light intensity detected by PD2 at time $t$ is
\begin{equation}
I_{c}(0 \mid d,v, \theta_{0}, \theta_{1},t)=|\alpha|^{2}-I_{c}(1 \mid d,v, \theta_{0}, \theta_{1},t).
\end{equation}
These average light intensities correspond to the beat signals of classical FMCW Lidar. This is the same as the beat signals obtained with $\nu=|\alpha|^{2}$ number of separable photons as in Eq. (\ref{equ12}) and (\ref{equ13}). This is expected as the average photon number in the coherent state is $|\alpha|^{2}$.
The resolutions of $\{d,v\}$ in this case are
\begin{equation}
\Delta d=2 \pi  \frac{c}{2 \Delta \omega}, \quad \Delta v=2 \pi \frac{c}{\omega_c  T^{r}_{m} },
\end{equation}
when one performs the DTF independently on the beat signal during each half period.

The fluctuation of light intensity of FMCW coherent state obeys the Poisson distribution. The probability to detect $x_{1}$ photons at PD1 is
\begin{equation}
p(x_{1}\mid \vec{x},t)=e^{-I_{c}(1 \mid \vec{x},t)}\frac{I_{c}(1 \mid \vec{x},t)^{x_{1}}}{x_{1} !} 
\end{equation}
and the probability to detect $x_{0}$ photons at at PD2 is
\begin{equation}
p(x_{0}\mid \vec{x},t)=e^{-I_{c}(0 \mid \vec{x},t)}\frac{I_{c}(0 \mid \vec{x},t)^{x_{0}}}{x_{0} !} ,
\end{equation}
The CFI matrix for each measurement is
\begin{equation}
F_{C}[\vec{x},t]_{i,j}=-\sum^{\infty}_{x_{0}=0} p(x_{0} \mid \vec{x},t)\frac{\partial^{2} \ln p(x_{0} \mid \vec{x},t)} {\partial x_{i} \partial x_{j} }-\sum^{\infty}_{x_{1}=0}p(x_{1} \mid \vec{x},t)\frac{\partial^{2} \ln p(x_{1} \mid \vec{x},t)} {\partial x_{i} \partial x_{j} }.
\end{equation}
Again, due to the additivity of CFI matrix $F_{C}[\vec{x},t]$ for each time $t$ of the measurement, the CFI matrix with the total detection time $T^{r}_{m}$ is
\begin{equation}
\begin{aligned}
F_{C}\left[d,v,\theta_{0},\theta_{1}\right]&=\int^{t_{d_{0}}}_{t_{d_{0}}-T^{r}_{m}/2}dt F_{C}\left[d,v,\theta_{0},\theta_{1},t\right]+\int^{t_{d_{1}}+T^{r}_{m}/2}_{t_{d_{1}}} dt F_{C}\left[d,v,\theta_{0},\theta_{1},t\right].
\end{aligned}
\end{equation}
By choosing $t_{d_{0}}=-t_{d_{1}}=T^{r}_{m}/4$ as in Sec. \ref{IIIB}, the classical CRB is
\begin{equation}
\begin{aligned}
\operatorname{Cov}(\left\{d,v,\theta_{0},\theta_{1}\right\})\geq  \frac{1}{|\alpha|^{2} T^{r}_{m}}
\begin{pmatrix}
 \frac{3c^{2}}{\Delta \omega^{2}}  & 0 & 0 & 0 \\ \\
0 &  \frac{12c^{2}}{\omega_{c}^{2} T_m^{2}}  & 0 & 0 \\ \\
0 & 0 & 2  & 0 \\ \\
0 & 0 & 0 &  2
\end{pmatrix},
\end{aligned}
\end{equation}
which is the same as Eq.(\ref{equ25}) with $|\alpha|^{2}=\nu$.

\section{The preparation of FMCW biphoton entangled state $|\psi_{2}\rangle$}

The FMCW biphoton entangled state $|\psi_{2}\rangle$ with triangular frequency modulation can be prepared through a nonlinear optical process with frequency modulation. To begin, unmodulated anti-frequency-correlated photon pairs are generated through spontaneous parametric down conversion (SPDC)\cite{jinOnChipGenerationManipulation2014,xueUltrabrightMultiplexedEnergyTimeEntangled2021}. These biphotons are subsequently modulated independently using optical phase modulators (OPMs) driven by the same linear frequency-sweeping (LFS) radio frequency (RF)\cite{luRealtimeThreedimensionalCoherent2020} with a center frequency of $\omega_r$ and modulation bandwidth of $\Delta\omega$, where $\omega_r$ exceeds $\Delta\omega$. Electro-optic modulators are selected as the OPMs for this process.

An obstacle in this process is the occurrence of several sidebands that are inherent when LSF-RF signals modulate light \cite{luRealtimeThreedimensionalCoherent2020}. These unwanted sidebands are separated by a frequency  of $\omega_r$ and have a bandwidth of $k\Delta\omega$ for the k-th sideband. The sidebands can be separated when the correlation bandwidth of the photon pair satisfies $3\sigma<\omega_r/2$, following the three-sigma guidelines. Given that the photon pairs are modulated independently, the undesired sideband of each photon can be filtered out using a narrow-band frequency filter, such as a grating fiber filter, etc.

\begin{figure}[h]
\centering
\includegraphics[width=0.7\linewidth]{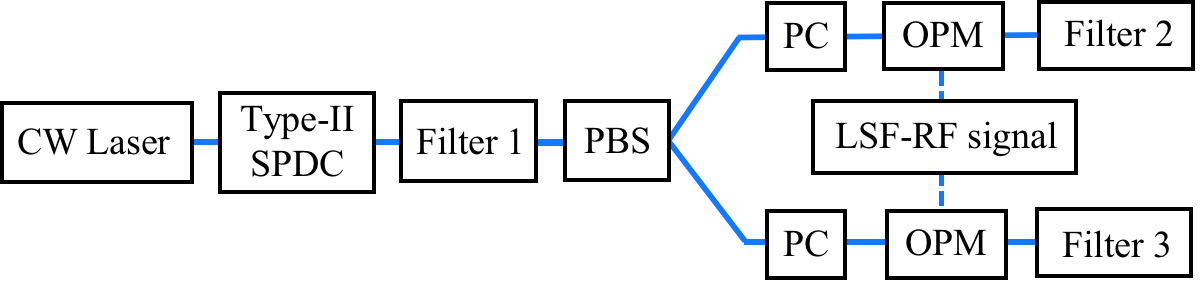}
\caption{A sketch for preparing the state $|\psi_{2}\rangle$ in the realm $3\sigma>\omega_r/2$. The central frequency of the Filter 1 matches the central frequency of the photon pair, and the central frequency of the Filter 2 and Filter 3 match the desired sideband. The bandwidth of three filters is the same and should be less than $3\sigma$.
The polarization beam splitter (PBS) is used to distribute the photon pair into two path. After the distribution, the polarization of each photon is rotated to the same direction and matches the requirement of OPM, by using polarization controller (PC).}
\label{fig:5}
\end{figure}

If $3\sigma>\omega_r/2$, there is a risk that the sidebands of each photon will overlap. In the most extreme case where $\sigma>>\omega_r$, the sidebands would be completely inseparable \cite{olislagerFrequencybinEntangledPhotons2010}. To address this challenge, we present two possible designs. The first solution is to filter the photon pairs before frequency modulation to ensure the separable sideband requirement. This is illustrated in Figure \ref{fig:5}. However, this comes at a cost of sacrificing a portion of the generated photon pairs, and the loss is more significant in the extreme case where $\sigma>>\omega_r$ \cite{olislagerFrequencybinEntangledPhotons2010}. In the second design, instead of filtering the photon pairs, unmodulated photons are fed into an optical circuit where sidebands counteract each other in a way that photon pairs within the desired sideband with minimal high-order sideband are produced from specific output ports. This design is an extension of I-Q modulator using in classical FMCW LiDAR \cite{gao2012frequency}.

It is worth mentioning that all the above preparation processes can be implemented using thin-film lithium niobate \cite{zhuIntegratedPhotonicsThinfilm2021} and are suitable for on-chip integration \cite{jinOnChipGenerationManipulation2014}.

\section{The nonlinear optical processes used in the quantum-enhanced pulsed LiDAR}

The capacity to estimate range and velocity simultaneously through quantum pulsed compression has been proven by quantum-enhanced pulsed LiDAR employing entangled pulsed light with a high time-bandwidth product \cite{shapiroQuantumPulseCompression2007}. However, further nonlinear optical processes are necessary to de-entangle the entangled pulsed light in order to measure the Time-of-Flight (ToF) and Doppler shift independently. In some quantum pulsed LiDAR proposals, these processes are also employed to entangle several pulsed light types to produce entangled pulsed light with high time-bandwidth product.

\begin{figure}[h]
\centering
\includegraphics[width=0.7\linewidth]{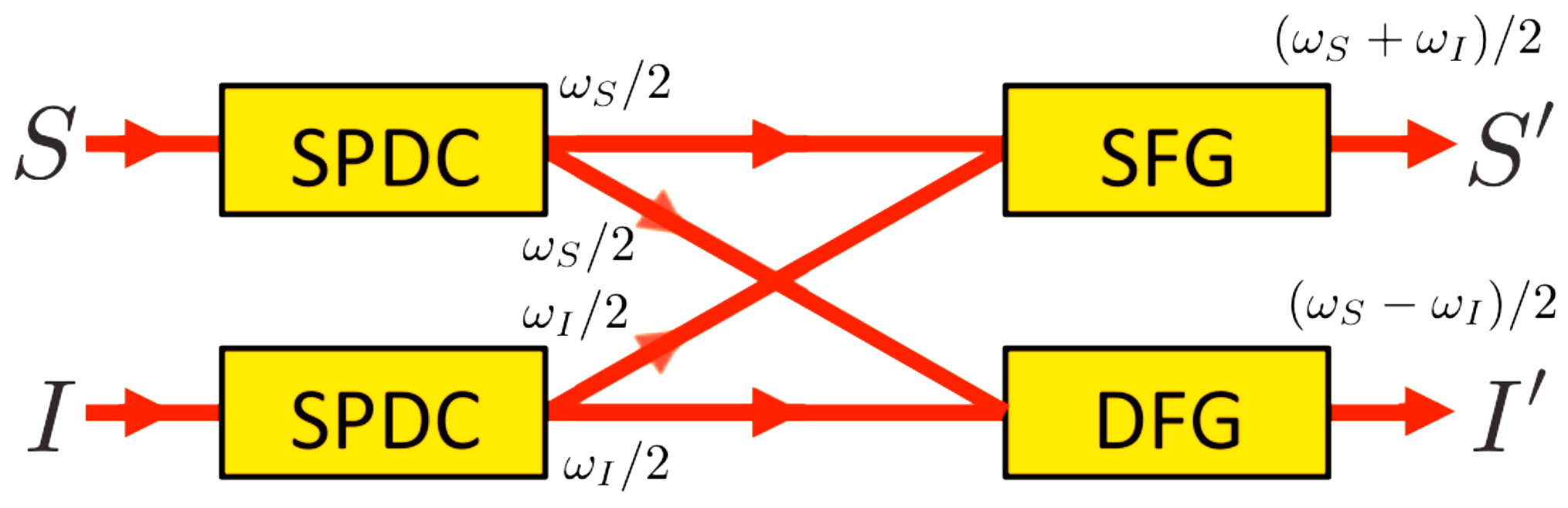}
\caption{The nonlinear optical processes used in the proposal of Ref. \cite{zhuangEntanglementenhancedLidarsSimultaneous2017}. SPDC: Single-photon-sensitive spontaneous parametric downconverter with extended phase matching. SFG: Single-photon-sensitive sum-frequency generator. DFG: Single-photon-sensitive difference-frequency generator.}
\label{fig:6}
\end{figure}

Fig. \ref{fig:6} illustrates the nonlinear optical processes used in the quantum-enhanced pulsed LiDAR proposal in Ref. \cite{zhuangEntanglementenhancedLidarsSimultaneous2017}. These processes enable the entanglement of two single-photon pulsed lights. Initially, the signal (idle) single-photon pulsed light, with a center angular frequency of $\omega_{S}$ $(\omega_{I})$, is split into two down-converted single-photon pulsed lights, each with a center angular frequency of $\omega_{S}/2$ $(\omega_{I}/2)$, through the use of spontaneous parametric down-convert(SPDC). Subsequently, both the signal and idle down-converted single-photon pulsed lights are combined using sum-frequency generator(SFG)/difference-frequency generator(DFG) to produce sum-frequency (difference-frequency) single-photon pulsed lights with a center angular frequency of $(\omega_{S}\pm\omega_{I})/2$. As the two down-converted single-photon pulsed lights are inherently frequency-entangled due to the nature of SPDC, the resulting sum-frequency single-photon pulsed light becomes entangled with the difference-frequency single-photon pulsed light. The processes illustrated in Fig. \ref{fig:6} can also be utilized in reverse to de-entangle the entangled pulsed light and are also utilized to construct the optimal measurement of time and frequency estimation for the proposal outlined in Ref. \cite{huangQuantumLimitedEstimationRange2021}.

The effectiveness of the aforementioned nonlinear optical processes, however, is constrained by the device's $\chi^{(2)}$ nonlinearity, and they are often less efficient than the linear optical processes used in the quantum FMCW LiDAR proposed here.

\bibliography{bibtex}